%% file: all_arxiv.tex
\documentclass[reprint,nofootinbib,superscriptaddress,aps]{revtex4-2}

\usepackage{placeins}
\usepackage{graphicx} % Required for inserting images
\usepackage{dcolumn}
\usepackage{xr}
\usepackage{bm}
\usepackage{wrapfig,lipsum,booktabs}
\usepackage[utf8]{inputenc} % allow utf-8 input
\usepackage[T1]{fontenc}    % use 8-bit T1 fonts
\usepackage{hyperref}       % hyperlinks
\usepackage{url}            % simple URL typesetting
\usepackage{booktabs}       % professional-quality tables
\usepackage{amsmath}
\usepackage{amsfonts} % blackboard math symbols
\usepackage{amssymb}
\usepackage{nicefrac}       % compact symbols for 1/2, etc.
\usepackage{xcolor}         % colors
\usepackage[capitalise]{cleveref}

\input{authors.tex}

\begin{document}

%\title{Active Liquid Crystal Theory Explains the Emergent Morphology and Collective Fluctuations of Human Mitotic Spindles}
\title{Active Liquid Crystal Theory Explains the Collective Organization of Microtubules in Human Mitotic Spindles}

\date{\today}

\begin{abstract}
\input{src/abstract.tex}
\end{abstract}

\RevtexAuthorBlock
\maketitle

\input{src/main.tex}

\bibliographystyle{unsrt}
\bibliography{references}

\onecolumngrid

% Title page for S.I.
\newpage
\thispagestyle{empty}
\begin{center}
  \vspace*{2cm}
  {\LARGE\bfseries Supporting Information for {\itshape Modified Active Nematic Model of
      HeLa Spindles}} \\[2.5em]
  % pull in the plain‐text author/affiliation block for SI
  \SIAuthBlock\\[3em]
  {\normalsize \today}
  \vfill
\end{center}
\newpage

% Reset equation, figure, and table numbers for the SI
\renewcommand{\thefigure}{S\arabic{figure}}
\setcounter{figure}{0}

\renewcommand{\theequation}{S\arabic{equation}}
\setcounter{equation}{0}

\renewcommand{\thetable}{S\arabic{table}}
\setcounter{table}{0}

% Reset sections but do not re-label
\setcounter{section}{0}
\setcounter{subsection}{0}

% \bibliographystyle{arxiv}
% \bibliography{references}

\input{src/supplementary}

\end{document}

%% file: authors.tex
\newcommand{\RevtexAuthorBlock}{%
\newcommand{\CorrespondenceNote}{Corresponding authors: \href{mailto:cokelleh@syr.edu}{cokelleh@syr.edu}, \href{mailto:smaddu@flatironinstitute.org}{smaddu@flatironinstitute.org}}%
  \author{Colm P. Kelleher}%
    \altaffiliation{These authors contributed equally to this work}%
    \thanks{\CorrespondenceNote}%
    \affiliation{Department of Molecular and Cellular Biology, Harvard University, Cambridge, MA 02138, USA}%
    \affiliation{Department of Physics, Syracuse University, Syracuse, NY 13244, USA}%
  \author{Suryanarayana Maddu}%
    \altaffiliation{These authors contributed equally to this work}%
    \thanks{\CorrespondenceNote}%
    \affiliation{Center for Computational Biology, Flatiron Institute, New York, NY, USA, 10010}
  \author{Mustafa Basaran}%
    \affiliation{Department of Molecular and Cellular Biology, Harvard University, Cambridge, MA 02138, USA}%
      \affiliation{John A. Paulson School of Engineering and Applied Sciences, Harvard University, Cambridge, MA 02138, USA}%
  \author{Thomas M{\"u}ller-Reichert}%
    \affiliation{Core Facility Cellular Imaging, Faculty of Medicine Carl Gustav Carus, \\Technische Universit{\"a}t Dresden, 01307 Dresden, Germany}%
  \author{Michael J. Shelley}%
    \affiliation{Center for Computational Biology, Flatiron Institute, New York, NY, USA, 10010}%
    \affiliation{Courant Institute of Mathematical Sciences, New York University, New York, NY, USA, 10012}%
  \author{Daniel J. Needleman}%
    \affiliation{Department of Molecular and Cellular Biology, Harvard University, Cambridge, MA 02138, USA}%
    \affiliation{John A. Paulson School of Engineering and Applied Sciences, Harvard University, Cambridge, MA 02138, USA}%
    \affiliation{Center for Computational Biology, Flatiron Institute, New York, NY, USA, 10010}%
}

\newcommand{\SIAuthBlock}{%
  Colm\,P.\,Kelleher\textsuperscript{1,2},\;
  Suryanarayana\,Maddu\textsuperscript{3},\;
  Mustafa\,Basaran\textsuperscript{1},\;
  Thomas\,M{\"u}ller-Reichert\textsuperscript{4},\\
  Michael\,J.\,Shelley\textsuperscript{3,5},\;
  Daniel\,J.\,Needleman\textsuperscript{1,6}
  \\[1ex]
  \textsuperscript{1}Department of Molecular and Cellular Biology, Harvard University, Cambridge, MA 02138, USA\\
  \textsuperscript{2}Department of Physics, Syracuse University, Syracuse, NY 13244, USA\\
  \textsuperscript{3}Center for Computational Biology, Flatiron Institute, New York, NY, USA, 10010\\
  \textsuperscript{4}Core Facility Cellular Imaging, Faculty of Medicine Carl Gustav Carus,\\ Technische Universit{\"a}t Dresden, 01307 Dresden, Germany\\
  \textsuperscript{5}Courant Institute of Mathematical Sciences, New York University, New York, NY, USA, 10012\\
  \textsuperscript{6}John A. Paulson School of Engineering and Applied Sciences, Harvard University, Cambridge, MA 02138, USA
}

%% file: src/abstract.tex
How thousands of microtubules and molecular motors self-organize into spindles remains poorly understood. By combining static, nanometer-resolution, large-scale electron tomography reconstructions and dynamic, optical-resolution, polarized light microscopy, we test an active liquid crystal continuum model of mitotic spindles in human tissue culture cells. 
The predictions of this coarse-grained theory quantitatively agree with the experimentally measured spindle morphology and fluctuation spectra. These findings argue that local interactions and polymerization produce collective alignment, diffusive-like motion, and polar transport which govern the behaviors of the  spindle's microtubule network, and provide a means to measure the spindle's material properties.
This work demonstrates that a coarse-grained theory featuring measurable, physically-interpretable parameters can quantitatively describe the mechanical behavior and self-organization of human mitotic spindles.

%% file: src/main.tex
\newcommand{\um}{\, \mu\text{m}}
\newcommand{\Rpm}{\mathbf{R}^{p}_{m}}
\newcommand{\Rim}{\mathbf{R}^{i}_{m}}
\newcommand{\Nim}{\widehat{\mathbf{N}}^{i}_{m}}
\newcommand{\Npm}{\widehat{\mathbf{N}}^{p}_{m}}
\newcommand{\Rvec}{\mathbf{R}}
\newcommand{\Qvec}{\mathbf{Q}}
\newcommand{\qvec}{\mathbf{q}}
\newcommand{\rvec}{\mathbf{r}}
\newcommand{\Nhat}{\widehat{\mathbf{N}}}
\newcommand{\nhat}{\widehat{\mathbf{n}}}
\newcommand{\NhatZ}{\widehat{\mathbf{N}}_{0}}
\newcommand{\dN}{\delta \mathbf{N}}
\newcommand{\dNy}{\delta N_{y}}
\newcommand{\dNz}{\delta N_{z}}
\newcommand{\dn}{\delta \mathbf{n}}
\newcommand{\NMT}{N_{\text{MT}}}
\newcommand{\Nseg}{N_{\text{seg}}}
\newcommand{\dNT}{\widetilde{\delta \mathbf{N}}}
\newcommand{\SNN}{S_{\text{NN}}}
\newcommand{\SCC}{S_{CC}}
\newcommand{\RCC}{R_{CC}}
\newcommand{\SNC}{S_{N_{y}C}}
\newcommand{\CNN}{C_{\text{NN}}}
\newcommand{\CCC}{C_{CC}}
\newcommand{\CNC}{C_{N_{y}C}}
\newcommand{\snn}{s_{\text{nn}}}
\newcommand{\cnn}{c_{\text{nn}}}
\newcommand{\cnc}{c_{\text{n}c}}
\newcommand{\Rcc}{R_{cc}}
\newcommand{\scc}{s_{cc}}
\newcommand{\ccc}{c_{cc}}
\newcommand{\snc}{s_{\text{n}c}}
\newcommand{\xhat}{\hat{\mathbf{x}}}
\newcommand{\yhat}{\hat{\mathbf{y}}}
\newcommand{\zhat}{\hat{\mathbf{z}}}
\newcommand{\dnyT}{\widetilde{\delta n}_{y}}
\newcommand{\dcT}{\widetilde{\delta c}}
\newcommand{\dCT}{\widetilde{\delta C}}
\newcommand{\dNyT}{\widetilde{\delta N}_{y}}
\newcommand{\dNzT}{\widetilde{\delta N}_{z}}
\newcommand{\dd}{\text{d}}
\newcommand{\w}{\omega}
\newcommand{\Pim}{\mathbf{P}^{i}_{m}}
\newcommand{\PimA}{\mathbf{P}^{i}_{m+1}}
\newcommand{\shat}{\mathbf{\hat{s}}}
\newcommand{\Phihat}{\mathbf{\hat{\Phi}}}
\newcommand{\nim}{\mathbf{n}^{i}_{m}}
\newcommand{\SN}{S^{\text{N}}}
\newcommand{\SC}{S^{c}}
\newcommand{\SNZero}{S_{0}^{\text{N}}}
\newcommand{\SCZero}{S_{0}^{C}}
\newcommand{\snZero}{s_{0}^{\text{n}}}
\newcommand{\scZero}{s_{0}^{c}}
\newcommand{\Sbe}{S^{\beta}}
\newcommand{\SalZero}{S_{0}^{\alpha}}
\newcommand{\SNyT}{\tilde{S}_y^\text{N}} 
\newcommand{\SNzT}{\tilde{S}_z^\text{N}}
\newcommand{\SCT}{\tilde{S}^C}
\newcommand{\SNT}{\tilde{S}^\text{N}}

The spindle is a highly dynamic cellular organelle that segregates chromosomes during cell division. It consists of rapidly polymerizing and depolymerizing microtubules and a diverse array of motor and non-motor microtubule-associated proteins (MAPs) \cite{karsenti2001mitotic}. This complex assembly provides a scaffold for force generation, ensuring high-fidelity chromosome segregation \cite{dumont2009force}. While the molecular composition of spindles is now relatively well-characterized \cite{neumann2010phenotypic, sauer2005proteome, wuhr2014deep}, the mechanisms driving its large-scale self-organization remain poorly understood.

\indent One approach to studying spindle self-organization involves developing coarse-grained models, similar to those traditionally used in materials physics \cite{chaikin1995principles}. In particular, continuum active hydrodynamic theories have emerged as powerful tools for studying cytoskeletal systems in which filaments are transiently coupled by a large number of motors and passive cross-linkers, which bind and unbind on timescales that are fast compared to the long-time, large-scale dynamics of the system. These theories have been instrumental in explaining the results of a series of experiments on a variety of microtubule-based systems, including purified \textit{in vitro} mixtures in which cytoskeletal filaments and motors spontaneously self-organize into dynamic ordered structures \cite{kruse2004asters, ramaswamy2010mechanics, saintillan2008instabilities}, \textit{Xenopus laevis} egg extract spindles \cite{brugues2014physical, oriola2018physics}, and spindles in intact mammalian cells \cite{kelleher2024long, conway2022self}. Of these systems, \textit{Xenopus} extract spindles are the best-characterized, in part because their large sizes $(\sim 40 \um)$ and long lifetimes ($\sim$ hours) facilitate detailed measurements of material properties, using either force probes or light microscopy techniques \cite{brugues2014physical, shimamoto2011insights}. In contrast, mammalian mitotic spindles are smaller ($\sim 10 \um$) and shorter-lived (with a well-defined metaphase steady-state lasting $\sim 10$ minutes \cite{chakraborty2008nucleoporin}), making them particularly challenging to study. As a result, much of our understanding of spindle material physics comes from studies of \textit{Xenopus} extract spindles.

However, recent advances in serial-section electron tomography have enabled full three-dimensional reconstructions of the positions and configurations of every microtubule in metaphase spindles in human tissue culture cells at a single instant in time \cite{kiewisz2022three, zimyanin2024using}. Separately, non-invasive polarization microscopy provides access to the time dynamics of the microtubule network in living spindles, but at a spatial resolution limited by the wavelength of visible light, $\sim 500$ nm. By combining data obtained using these complementary techniques, we probe the physical basis of microtubule organization in HeLa metaphase spindles, and demonstrate that spatiotemporal patterns in microtubule orientations and densities are predicted by a minimal coarse-grained model based on the physics of active liquid crystals \cite{marchetti2013hydrodynamics}. This work indicates that the organization of microtubules in spindles is driven by their turnover dynamics and collective co-alignment, diffusion, and polar transport. Furthermore, the model enables inference of key organelle-scale material properties, including parameters describing nematic elasticity and microtubule diffusivity, and quantitatively reproduces the statistics of the arrangement of microtubules in cross section at the metaphase plate without additional fitting parameters. Taken together, this work demonstrates that coarse-grained, active-liquid-crystal-based models provide a quantitative framework for investigating microtubule self-organization and inferring key organelle-scale properties \textit{in vivo}.

\section*{Results}
\begin{figure*}[t!]
  \includegraphics[width=\textwidth]{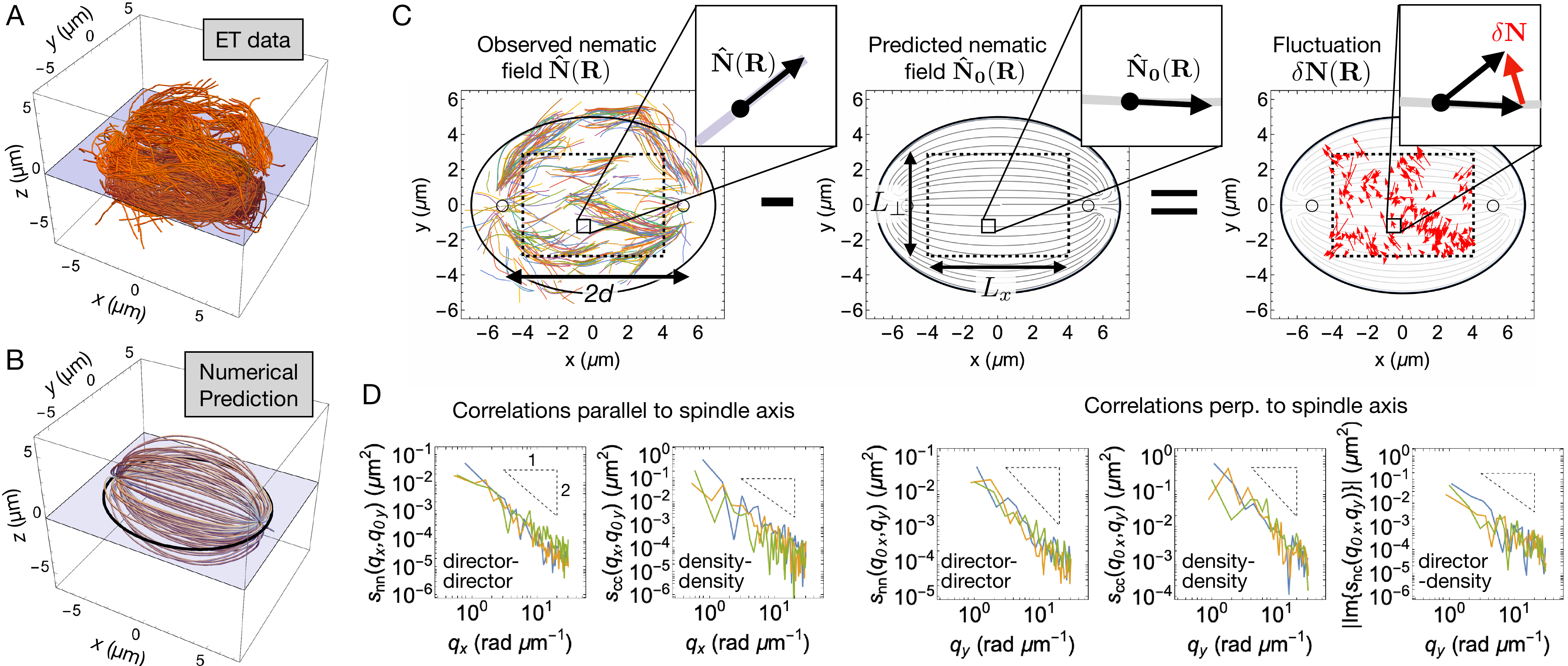}
  \caption{Spatial correlations in fluctuations calculated from electron tomography reconstructions of HeLa spindles.
 (A) Electron tomography enables complete reconstruction of HeLa spindle microtubules.
Only microtubules with contour length $ >1 \um$ are displayed here, but all microtubules are used in
subsequent calculations. The light blue plane indicates $z = 0$.
(B) Active liquid crystal model of microtubule orientation. Grey curves show streamlines of the numerically predicted director field $\Nhat_{0}(\Rvec)$; black curve in $z = 0$ plane
indicates the outline of the ellipsoid that best fits the boundary of the spindle shown in A.
(C) At each microtubule segment, the director fluctuation $\dN (\Rvec)$ (right) is calculated by subtracting the predicted director $\Nhat_{0}(\Rvec)$ (middle) from the observed director $\Nhat (\Rvec)$ (left). In the $\Nhat(\Rvec)$ plot, colored curves show microtubules that lie within $0.5 \um$ of the plane $z=0$. In all plots, black circles indicate centriole positions; black dashed box shows the region (dimensions $L_x \times L_{\perp} \times L_{\perp}$) where correlations are calculated.
(D) Director-director, density-density, and director-density correlation functions $\snn (\qvec)$, $\scc(\qvec)$, and $\snc(\qvec)$ along wave-vector components $q_{x}$ and $q_{y}$, parallel and perpendicular to the spindle axis respectively, for $n_{\text{ET}}=3$ different spindles. The cross-correlation $\snc(\qvec)$ is shown as a function of $q_{y}$ only (rightmost plot). For all curves plotted as a function of $q_{y}$, correlation functions were averaged over 10 rotations of the spindle about its central axis $\xhat$. Dashed black triangles in all plots indicate slope -2, corresponding to a functional form $\propto q^{-2}$.}
\label{Fig1}
\end{figure*}

\indent \textbf{Microtubule Density and Orientation Fields Estimated from Electron Tomography Data Are Consistent with Active Liquid Crystal Theory.}
Electron tomography provides detailed ultrastructural information on microtubules and chromosomes within spindles, allowing for precise measurements of each microtubule’s length, position, and morphology within the spindle apparatus (\cite{muller2018mitotic, zimyanin2024using}, Fig.~\ref{Fig1}A). We used recent electron tomography reconstructions of HeLa spindles \cite{kiewisz2022three} to estimate continuous orientation and density fields in the spindle bulk. To do this, we divided microtubule $p$ into $M_{p}$ sub-segments, with each sub-segment $m$ characterized by its three-dimensional (3D) position $\Rpm$, and 3D orientation vector $\Npm$, both determined from the endpoint locations of the segment (S.I.~1). In the three complete electron tomography reconstructions we analyzed, the total number of microtubules ($\NMT$) varies between 4000 and 7300, with an average of approximately 110 sub-segments per microtubule. The coarse-grained fields describing the microtubule orientation $\Nhat(\Rvec)$ and density $\rho(\Rvec)$ at position $\Rvec$ are estimated from the positional and orientational information of the sub-segments ~\cite{chaikin1995principles},
\begin{equation}
\rho(\Rvec) = \sum_{p=1}^{\NMT} \sum_{m=1}^{M_{p}}\delta(\Rvec -\Rpm)
\end{equation}
\begin{equation}
\Nhat(\Rvec) = \frac{V}{\Nseg}\sum_{p=1}^{\NMT} \sum_{m=1}^{M_{p}}\Npm \delta(\Rvec -\Rpm)
\end{equation}
where $V$ is the analysis volume, $\Nseg = \sum_{p}M_{p}$ is the total number of microtubule segments in the volume, and the nematic director $\Nhat(\Rvec)$ is normalized to unit length at each point $\mathbf{R}$.
%\mathbf{N(R)} = \frac{V}{\Nseg}\sum_{p=1}^{\NMT} \sum_{m=1}^{M_{p}}\Npm \delta(\Rvec -\Rpm), \:\: \Nhat(\Rvec) = \frac{\mathbf{N(R)}}{\Vert \mathbf{N(R)}\Vert},

To interpret these coarse-grained estimates, we compared them to predictions from an active liquid crystal theory of dynamic microtubules collectively interacting via cross-linkers and molecular motors. We use a theory in which the orientation of microtubules is determined by nematic interactions, with a polar transport of microtubules aligned along the nematic director. Neglecting random noise terms and considering a state without hydrodynamic flows, as has previously been argued is appropriate for metaphase spindles, this theory predicts that the fields $\rho(\Rvec,t)$ and $\Nhat(\Rvec,t)$ obey the equations of motion (\cite{brugues2014physical}, S.I.~3)
\begin{align}
\frac{\partial \rho}{\partial t}  &= \Gamma_0 - \Theta \rho + \nabla \cdot \left[ D \nabla \rho - v_{1} \rho \Nhat \right] \label{eq:mt_density}\\
%\frac{\partial N_\alpha}{\partial t} &= K \nabla^{2} N_{\alpha}. 
\frac{\partial \Nhat}{\partial t} &= K (\mathbf{I} - \Nhat \Nhat)\nabla^{2} \Nhat. 
\label{eq:mt_director}
\end{align}

\noindent In the equation for microtubule density $\rho$, the first term represents microtubule nucleation with rate $\Gamma_0$. The second term describes microtubule disassembly events (catastrophes) occurring at rate $\Theta$. The term inside the divergence operator captures spatial transport of microtubules by both diffusive-like motion and directed (polar) transport. Microtubule diffusive-like motion is characterized by a diffusivity $D$, while directed (polar) transport follows the vector field $v_{1} \Nhat$. The latter term explicitly violates nematic symmetry and is uniquely active in origin, since it breaks detailed balance~\cite{marchetti2013hydrodynamics}. The equation for the director results from using the one-Frank constant approximation with parameter $K$, the nematic diffusivity, given by the ratio of the nematic elastic constant to the rotational drag coefficient~\cite{brugues2014physical, kelleher2024long}. The projection operator $\mathbf{I} - \Nhat \Nhat$ preserves the unit magnitude of the orientation field. To the lowest order in spatial gradients, the theory predicts that steady-state microtubule density $\rho_{0}$ is constant in space and time, while the steady-state director $\Nhat_{0}$ obeys the projected vectorial Laplace equation,
\begin{equation}
\rho_{0}  = \frac{\Gamma_0}{\Theta}; \qquad 
\left(\mathbf{I - \Nhat_{0} \Nhat_{0} }\right)\nabla^{2} \Nhat_{0} = 0. \label{eq:NhatSS}
\end{equation}

It was previously shown that Eq.~(\ref{eq:NhatSS}) for the director $\Nhat$, when combined with anchoring conditions derived from the geometry of the best-fit spindle boundary, accurately describes the average microtubule orientation field in HeLa cells~\cite{conway2022self}. This result agrees quantitatively with our independent analysis of microtubule orientation patterns across the spindle (Figs.~1B \& S1). Additionally, within the spindle interior, the coarse-grained microtubule density is approximately uniform, varying by at most $\sim 20\%$ near the metaphase plate, where condensed chromosomes occupy a significant fraction of the available volume (Fig.~S2). Taken together, the spindle-scale patterns of microtubule concentration and orientation quantitatively agree with the theory.\\

\indent \textbf{Properties of Density and Orientation Correlation Functions from Electron Tomography Data.}
\indent We next investigated spatial patterns in fluctuations of microtubule density and orientation. These patterns, as quantified by correlation functions, provide a validation of the equations of motion Eq.~\eqref{eq:mt_density} and Eq.~\eqref{eq:mt_director}, and allow estimation of the physically meaningful parameters in those equations. To remove the dependence of $\rho$ on the density of microtubule sub-segments, which depends on experimental details such as sampling density, we instead analyze the normalized density, $C = \rho/\rho_{0} =  \rho V/\Nseg$. Here, $V$ is the volume of the analysis box, defined as the largest-volume cuboid contained within the spindle boundary; the dimensions of $V$ are $L_{x} \times L_{\perp} \times L_{\perp}$ (Fig.~1C, black dashed box). This normalization procedure also facilitates direct comparison with polarization microscopy measurements (next section). Fluctuations in normalized density are defined as
\begin{equation}\label{eq:c_fluctuation}
\delta C(\Rvec) = \frac{V}{\Nseg}\sum_{p=1}^{\NMT} \sum_{m=1}^{M_{p}}\delta(\Rvec -\Rpm)- 1.
\end{equation}
Fluctuations in orientation are calculated by subtracting the director field predicted by Eqn.~\eqref{eq:NhatSS} from the observed director field (Fig.~1C), 
\begin{equation}\label{eq:n_fluctuation}
  \dN(\Rvec) = \Nhat(\mathbf{R}) - \NhatZ(\Rvec) .
\end{equation}
To quantify spatial patterns in fluctuations of microtubule concentration and orientation fields, we plot equal time correlation functions, measuring how fluctuations at different spatial locations are related at the same instant in time (see \textcolor{blue}{Materials \& Methods}).
%To quantify spatial patterns in fluctuations in the microtubule concentration and orientation fields, we plot the equal-time density-density, director-director, and director-density correlation functions (\textcolor{blue}{Materials \& Methods}). 
We calculate density-density, director-director, and director-density correlation functions in Fourier space, where they are predicted to take relatively simple functional forms \cite{chaikin1995principles}, and plot them as a function of the wavevector components along the long and short axes of the spindle (Fig.~\ref{Fig1}D). To facilitate comparison with polarization microscopy measurements (following section), we project the fluctuating fields over the $\zhat$-axis and compute the correlation functions in the $xy$-plane (Materials \& Methods, S.I.~3). For each of the $n_{\text{ET}}=3$ electron tomography reconstructions, we first plot the director-director $\snn(q_{x},q_{0y})$ and density-density $\scc(q_{x},q_{0y})$ correlation function along the wave vector component $q_{x}$ parallel to the spindle axis. We plot these correlation functions in the long-wavelength limit, i.e.~with the perpendicular component $q_{y}$ fixed at the lowest available mode $q_{0y}=2\pi/L_{\perp}$ (\textcolor{blue}{Materials \& Methods}). We also plot $\snn$ and $\scc$ as a function of the wavevector component $q_{y}$ perpendicular to the spindle's axis, with the parallel component fixed at the lowest available mode $q_{0x}=2\pi/L_{x}$. All these correlation functions approximately follow inverse-square power-law scaling, $1/q^{2}$, consistent with a minimal model in which orientation and density dynamics are governed by independent local processes, for example, nematic elasticity and microtubule diffusive-like motion respectively (\cite{chaikin1995principles}, S.I.~3). However, we also observe that the director-density cross-correlation $\snc(q_{0x},q_{y})$ is finite (Fig.~\ref{Fig1}D, far right), with magnitude and functional form comparable to the self-correlations. This observation indicates that density and orientation dynamics are not independent and suggests that coupling between the fields significantly influences their behavior. \\

\begin{figure*}[t!]
  \includegraphics[width=\textwidth]{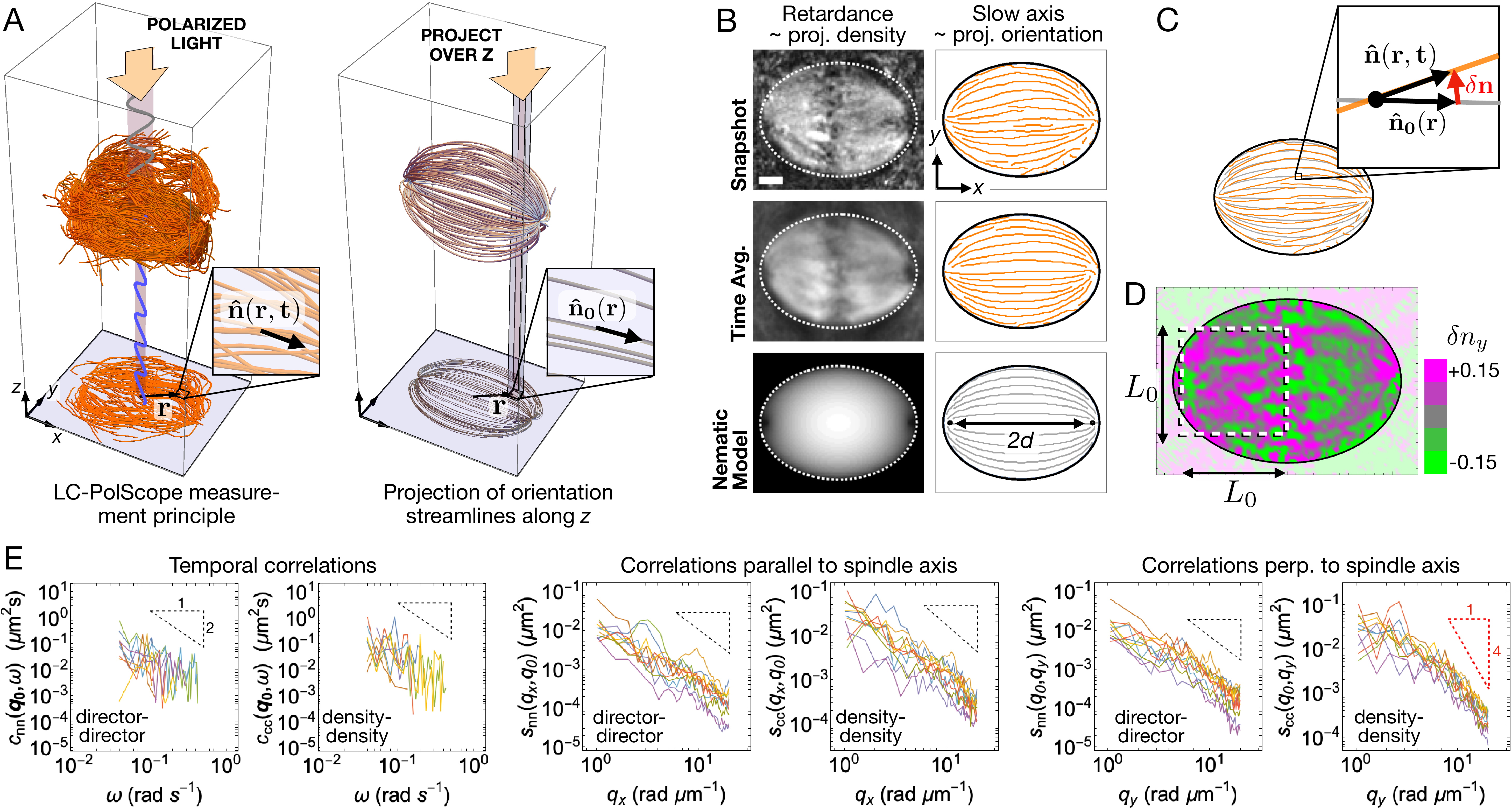}
  \caption{LC-PolScope characterization of spindles in living HeLa cells. (A) \textit{Left}: LC-PolScope images take advantage of the intrinsic birefringence of spindle microtubules to provide information about their density and orientation, projected over the optical axis $\zhat$. \textit{Right}: LC-PolScope images are compared to the predictions of a model where the 3D streamlines predicted by the nematic model are projected over $\zhat$ using Eq.~(\ref{eq:polscope_2d}). (B) \textit{Top \& Middle Rows}: Instantaneous and time-averaged images of LC-PolScope retardance (related to projected microtubule density) and optical slow axis (related to projected orientation). Time averages taken over 5 mins (34 frames). White dashed curve in the retardance images shows the best-fit ellipse. \textit{Bottom Row}: Predictions of retardance and slow axis from numerical solutions of Eqn.~\ref{eq:NhatSS} supplemented with boundary conditions based on the geometry of this spindle (S.I.~3). Defect positions (black dots) are chosen so that the predicted orientation field fits the observed time-averaged orientation field. Scale bar 2 $\um$. (C) 2D-projected director fluctuations $\dn (\rvec, t)$ are calculated by subtracting the numerically predicted 2D director $\nhat_{0}(\rvec)$ from the measured 2D director $\nhat(\rvec, t)$. (D) Instantaneous plot of $\delta n_y(\rvec, t)$, the $y$-component of $\dn (\rvec, t)$. Near the middle of the spindle, $\dn \approx \delta n_y \yhat$. Correlations are calculated in the white dashed box, dimensions $L_{0} \times L_{0} = 6 \um \times 6 \um$. (E) \textit{Left to right}: Spatiotemporal director-director and density-density correlation functions $\cnn(\qvec_0, \w)$ and $\ccc(\qvec_0, \w)$, evaluated at the longest available wave-vector $\qvec_0 = (2\pi/L_{0}, 2\pi/L_{0})$, as a function of frequency $\w$. The spatial correlation functions $\snn(\qvec)$ and $\scc(\qvec)$, plotted as functions of wave-vector components $q_{x}$ and $q_{y}$, parallel and perpendicular to the spindle axis respectively. Dashed black triangles in the first five plots indicate a slope of $-2$ ($\propto q^{-2}$ scaling); the bold red dashed triangle in the final plot indicates a slope of $-4$ ($\propto q^{-4}$ scaling). Each colored curve represents data from one spindle; data from $n_{\text{Pol}} = 11$ different spindles is shown.}
\label{Fig2}
\end{figure*}

\indent \textbf{Microtubule Density and Orientation Fields Estimated from LC-PolScope Data Are Consistent with Active Liquid Crystal Theory.}
Next, we sought to measure the time-resolved dynamics of microtubule density and orientation fluctuations to complement the spatial correlation measurements obtained from static electron tomography reconstructions. We used the LC-PolScope \cite{oldenbourg1996new,liu2000reliable}, a form of non-invasive, label-free polarized light microscopy, to characterize the structure and dynamics of the microtubule network in HeLa spindles. The LC-PolScope simultaneously measures the optical retardance $r(\rvec, t)$ and optical slow axis $\theta(\rvec, t)$ at a given time $t$ and each position $\rvec$ in a two-dimensional (2D) image \cite{liu2000reliable}. Although the LC-PolScope has lower spatial resolution than electron tomography (S.I.~2), it allows us to extract time-resolved information regarding the microtubule cross-sectional density $\rho_{2D}(\Rvec,t)$ and the director $\Nhat(\rvec,t)$ (\cite{kelleher2024long, conway2022self}. In general, the retardance value depends on the angle between the director and the optical axis $\zhat$ (S.I.~2). However, if microtubules are perpendicular to the optical axis $\zhat$, $\Nhat \cdot \zhat \approx 0$, as is the case for microtubules in the central spindle region (Fig.~\ref{Fig2}A\&B),
\begin{gather}
   r(\rvec, t) \approx A_0 \int_T \rho_{2D}(\Rvec, t) d z; \label{eq:retard}\\
   \hat{\mathbf{n}}(\rvec, t)  \equiv 
( \cos\theta(\rvec, t), \sin\theta(\rvec, t) ) 
\approx 
\frac{\int_T \Nhat (\Rvec, t) \, dz}{\left\|\int_T \Nhat(\Rvec, t) \, d z\right\|} \label{eq:polscope_2d},
\end{gather}
where $T$ is the sample thickness, $A_{0} \approx 7.5$ nm$^{2}$ is a previously-measured constant that characterizes the retardance contribution of a single microtubule~\cite{oldenbourg1998birefringence, kelleher2024long}, and $\rho_{2D}(\Rvec)$ is the 2D microtubule cross-sectional density in the plane perpendicular to $\Nhat(\Rvec)$. In the central spindle, microtubules are approximately parallel to the spindle axis $\xhat$, and $\rho_{2D}(\Rvec) = \lambda\rho(\Rvec)$, where $\lambda$ is the electron tomography sampling length along the microtubules, $\lambda \approx 20 $ nm~\cite{kiewisz2022three}. It was previously shown that retardance and slow axis measurements of HeLa spindles are quantitatively consistent both with electron tomography reconstructions and with the predictions of Eqn.~\ref{eq:NhatSS}~\cite{conway2022self}. This is also consistent with our independent measurements and analysis (Fig.~S5). \\

\indent \textbf{Properties of Density and Orientation Correlation Functions from LC-PolScope Data.}
We next sought to compare in more detail the structure and dynamics of the microtubule network, as measured by LC-PolScope, with the results of our analysis of electron tomography data. To perform this comparison, we computed the spatiotemporal correlation functions of director and density fluctuations from $n_{\text{Pol}}=11$ LC-PolScope datasets, focusing on several long-wavelength limits (Fig.~2C). To facilitate comparison with electron tomography data, we normalize the retardance measurements in each spindle by its time-averaged value inside the analysis box, $c(\rvec,t) = r(\rvec,t)/\langle r(\rvec) \rangle_{t}$  (\textcolor{blue}{Materials \& Methods}). To minimize interference from metaphase chromosomes, we analyzed fluctuations in an analysis box of dimensions $L_{0} \times L_{0} = 6 \um \times 6 \um$, centered at $(- 3 \um, 0 )$ (Fig.~2D). Fluctuations are defined via continuous two-dimensional analogs of Eqns.~\eqref{eq:c_fluctuation} and \eqref{eq:n_fluctuation},
 \begin{align*}
   \delta c(x,y) &= c(x,y) - 1; \\
   \dn (x,y) &= \nhat (x,y) - \nhat_{0}(x,y),
\end{align*}
  where $\nhat_{0}(x,y)$ is the orientation field predicted by Eqn.~\eqref{eq:NhatSS}, projected into 2D along a direction perpendicular to the spindle axis using Eq.~(\ref{eq:polscope_2d}). We observed that the autocorrelation functions, $\cnn(\qvec_{0}, \omega)$ and $\ccc(\qvec_{0}, \w)$, with wavenumbers fixed at the lowest nonzero modes $\qvec_{0} = ( q_{0}, q_{0}) = ( 2\pi/L_{0}, 2\pi /L_{0} )$ (Fig.~\ref{Fig2}E), decay roughly proportional to $1/\w^2$. At large $q$, the equal-time correlation functions $\snn (q_{x}, q_{0})$ and $\scc (q_{x}, q_{0})$ appear consistent with inverse-square power-law scaling, similar to the electron tomography measurements. At large $q_{y}$, however, the density correlation $\scc (q_0, q_y)$ displays scaling between $1/q_{y}^2$ and $1/q_{y}^4$, inconsistent with collective dynamics controlled by microtubule diffusion alone. At low wavenumbers (long wavelengths), all equal-time correlation functions calculated from LC-PolScope data deviate from power-law behavior and start to flatten; this trend is most clearly evident in $\scc(q_{0}, q_{y})$. This could be a consequence of the finite size of our analysis box, since power-law behavior is only expected in the infinite system-size limit. Alternatively, it could indicate that other processes, for instance, microtubule turnover, become dominant at the longest wavelengths.\\

  \indent \textbf{Average Orientational Data from Electron Tomography and LC-PolScope is Consistent with Active Liquid Crystal Theory.}
  We next sought to compare orientation data, simultaneously averaged over all electron tomography and all LC-PolScope data sets, with the predictions of Eqn.~\eqref{eq:NhatSS}. Averaging orientation data across multiple spindles is non-trivial because predictions from Eqn.~\eqref{eq:NhatSS} depend explicitly on the geometry of each individual spindle. In our model, spindle geometry is characterized by three independent parameters (S.I.~1): the best-fit spindle half-length $a$, best-fit spindle half-width $b$, and half-distance between spindle poles $d$. (The quantity $d$ is measured directly from electron tomography data, but is treated as fit parameter when determining the predicted orientation fields for LC-PolScope slow axis data). To average over spindles, we first rescaled the spatial coordinates $(x,y,z)$ of each dataset by its interpolar half-distance $d$, and then computed the average over all available datasets: 
\begin{equation*}
\langle \theta \rangle(x/d,s/d) = \langle \theta (x/d,s/d) \rangle_{\text{all spindles}}.
\end{equation*}
where \( s = \sqrt{y^2 + z^2} \) denotes the radial coordinate. When taking the average over spindles, we give equal weight to the $n_{\text{ET}}=3$ $\zhat$-projected electron tomography data sets and the $n_{\text{Pol}}=11 $ time-averaged LC-PolScope data sets. We compare this combined dataset to numerical simulations in which the spindle poles are positioned at $(\pm 1, 0, 0)$. In these simulations, we enforce tangential boundary conditions on an ellipsoid whose long and short semi-axes were set to the averages of the rescaled spindle dimensions $a/d$ and $b/d$, with averages again taken over all datasets (equal weights given to electron tomography and LC-PolScope). Without using additional fit parameters, the combined dataset shows excellent agreement with the numerical predictions (Fig.~3A,B).

\indent \textbf{Active Liquid Crystal Theory Quantitatively Explains Combined Electron Tomography and LC-PolScope Fluctuation Spectra}
We then investigated whether estimates of spatiotemporal correlation functions, obtained from electron tomography and LC-PolScope data, are consistent with the predictions of Eqns.~\eqref{eq:mt_density} and \eqref{eq:mt_director}. To do this, 
we first binned each data set and plotted them on a common set of axes (Fig.~3C). This reveals that the two estimates of $\snn(q_{x}, q_{0})$ and $\snn(q_{x}, q_{0})$ agree, within experimental uncertainty, for all values of $q_{x}$. Along $q_{y}$, the two estimates of the director-director correlation function agree, except at the very lowest values of $q_{y}$. The density-density correlation functions and the density-director correlation function have a similar shape between both data sets, but the LC-PolScope estimates were systematically lower. The close agreement between these independent measurements--achieved without using any adjustable parameters--is remarkable, and indicates that experimental errors are relatively minor, such as potential fixation and reconstruction artifacts in the electron tomography data and potential optical projection artifacts and camera noise in the LC-PolScope data. 

\indent Having established that LC-PolScope and electron tomography measurements yield similar correlation spectra, we next focused on using all available fluctuation data to test the active liquid crystal model and to estimate the phenomenological parameters it contains. To do this, we combined the spatial correlation function sets by directly averaging the LC-PolScope and electron tomography estimates, giving equal weight to each data set (Fig.~3C, black points with error bars). For the temporal correlation functions, only LC-PolScope data were available (Fig.~3C, green circles with error bars). To compare the combined correlation functions with the predictions of Eqs.~(\ref{eq:mt_density}) and (\ref{eq:mt_director}), we introduced noise terms into the equations of motion and calculated the resulting spatiotemporal density–density, $\ccc(\qvec,\w)$, and director-director, $\cnn (\qvec, \omega)$, correlations in Fourier space (S.I.~3),
\begin{align}
    \ccc(\qvec,\w)  &= \frac{q_{y}^2 v_{1}^{2} (\snZero)^2 + (K^2q^4+\w^2) (\scZero)^2}{(K^2q^4+\w^2)\left((q_{x} v_{1}+\w)^2+(D q^{2}+\Theta)^{2}\right)}; \nonumber \\
\cnn (\qvec, \omega)  &= \frac{ (\snZero)^{2} }{\omega^{2}  + K^{2} q^{4}},
\end{align}
and the equal-time density-density, $\scc(\qvec)$, director-director, $\snn (\qvec)$, and density-director, $\snc(\qvec)$, correlations in Fourier space,
\begin{gather}    
    \scc(\qvec) = \frac{(\scZero)^{2}}{2(D q^{2}+\Theta)}
+ \nonumber \\ \frac{q_{y}^{2}(\snZero)^{2}v_{1}^{2}((D+K)q^{2}+\Theta)}
{2Kq^{2}(Dq^{2}+\Theta)\bigl(q_{x}^{2}v_{1}^{2}+((D+K)q^{2}+\Theta)^{2}\bigr)}; \nonumber \\
\snn (\qvec) = \frac{(\snZero)^{2}}{2 K q^{2}}; \nonumber \\
     \snc(\qvec)  = -\frac{i q_y v_1 (\snZero)^{2}}{2 K q^{2} ((D+K)q^{2}+ i q_{x} v_{1} + \Theta) }.
\label{eq:FourierCFs}
\end{gather}

In these equations, $\scZero$ and $\snZero$ are the magnitudes of the Gaussian-distributed random noise in the density and orientation equations, respectively. All other parameters are defined in Eqns.~\eqref{eq:mt_density} and \eqref{eq:mt_director} and described in the text immediately following those equations. This model for fluctuations is similar to that presented in \cite{brugues2014physical}, with one important modification: whereas the model in that work only contained random noise in the orientation field, the model presented here includes independent sources of noise in both fields. This modification is needed to explain the magnitude and form of the density correlations $\scc(q_{x},q_{0y})$ in the direction parallel to the spindle axis (Fig.~3C, second column), and the cross-correlations between the director and density fields.\\

To test whether these predictions are consistent with the combined correlation functions, we estimated $\Theta = 0.047 s^{-1}$ from previous measurements in HeLa cells~\cite{conway2022self}, and used all remaining parameters ($\scZero$, $\snZero$, $D$, $K$, and $v_{1}$) as fit parameters. (When estimating $\Theta$, we used the average of the turnover frequencies of kinetochore and non-kinetochore microtubules, weighted to account for the relative abundance of these populations.) With the best-fit parameters given in Table~1, the model produces excellent fits to the observed data  (Fig.~3C, black curves). As in the previous analysis of fluctuations in \textit{Xenopus} egg extract spindles, we found that the best-fit microtubule diffusion constant, $D = (0.0043 \pm 0.0023) \um^{2}$ s$^{-1}$, is similar to the best-fit nematic diffusivity, $K = (0.0021 \pm 0.0002) \um^{2}$ s$^{-1}$ (uncertainties estimated via non-replacement subsampling, (\textcolor{blue}{Materials \& Methods})). The values we measured are lower than the \textit{Xenopus} values by a factor of approximately five \cite{brugues2014physical}, which might be caused by microtubules being shorter in HeLa spindles than in \textit{Xenopus} extract spindles, $\sim 2 \um$ and $\sim 6 \um$ respectively, or microtubules being less dense in HeLa spindles than in \textit{Xenopus} extract spindles, $\sim 13$ microtubules$/\um^{2}$ and $\sim 50$ microtubules$/\um^{2}$ in cross-section respectively (S.I.~1 \& 2, \cite{brugues2012nucleation}. The value we measured for $v_{1}$, $(8.4 \pm 6.2) \um$/min, is larger than the speed of microtubule sliding in these spindles, $\sim 1 \um$/min \cite{conway2022self}, but smaller than the speed of microtubule polymerization in these spindles, $\sim 15 \um$/min (data not shown). Such an intermediate polar transport speed could result from microtubule-mediated microtubule nucleation \cite{decker2018autocatalytic, petry2013branching, ishihara2016physical}. \\
%(CITE Decker F, Oriola D, Dalton B, Brugués J. Autocatalytic microtubule nucleation determines the size and mass of Xenopus laevis egg extract spindles. Elife. 2018 Jan 11;7:e31149. doi: 10.7554/eLife.31149. PMID: 29323637; PMCID: PMC5814149. CITE Petry S, Groen AC, Ishihara K, Mitchison TJ, Vale RD. Branching microtubule nucleation in Xenopus egg extracts mediated by augmin and TPX2. Cell. 2013 Feb 14;152(4):768-77. doi: 10.1016/j.cell.2012.12.044. PMID: 23415226; PMCID: PMC3680348. CITE Ishihara K, Korolev KS, Mitchison TJ. Physical basis of large microtubule aster growth. Elife. 2016 Nov 28;5:e19145. doi: 10.7554/eLife.19145. PMID: 27892852; PMCID: PMC5207775).\\

\textbf{Active Liquid Crystal Theory Explains The Arrangement of Microtubules in Spindle Cross-Sections.}

Microtubules in spindles, both kinetochore microtubules and non-kinetochore microtubules, are often described as being organized into bundles \cite{tolic2018mitotic,wadsworth2021multifunctional,valdez2023mechanisms,vukuvsic2017microtubule,kajtez2016overlap}.
%(CITE Tolić IM. Mitotic spindle: kinetochore fibers hold on tight to interpolar bundles. Eur Biophys J. 2018 Apr;47(3):191-203. doi: 10.1007/s00249-017-1244-4. Epub 2017 Jul 19. PMID: 28725997; PMCID: PMC5845649; Wadsworth P. The multifunctional spindle midzone in vertebrate cells at a glance. J Cell Sci. 2021 May 15;134(10):jcs250001. doi: 10.1242/jcs.250001. Epub 2021 May 27. PMID: 34042161; Valdez VA, Neahring L, Petry S, Dumont S. Mechanisms underlying spindle assembly and robustness. Nat Rev Mol Cell Biol. 2023 Aug;24(8):523-542. doi: 10.1038/s41580-023-00584-0. Epub 2023 Mar 28. PMID: 36977834; PMCID: PMC10642710; Vukušić K, Buđa R, Bosilj A, Milas A, Pavin N, Tolić IM. Microtubule Sliding within the Bridging Fiber Pushes Kinetochore Fibers Apart to Segregate Chromosomes. Dev Cell. 2017 Oct 9;43(1):11-23.e6. doi: 10.1016/j.devcel.2017.09.010. PMID: 29017027; PMCID: PMC5637169; Kajtez J, Solomatina A, Novak M, Polak B, Vukušić K, Rüdiger J, Cojoc G, Milas A, Šumanovac Šestak I, Risteski P, Tavano F, Klemm AH, Roscioli E, Welburn J, Cimini D, Glunčić M, Pavin N, Tolić IM. Overlap microtubules link sister k-fibres and balance the forces on bi-oriented kinetochores. Nat Commun. 2016 Jan 5;7:10298. doi: 10.1038/ncomms10298. PMID: 26728792; PMCID: PMC4728446).
The active liquid crystal theory presented here successfully describes aspects of microtubule organization in the spindle without explicitly invoking bundling. In this theory, a ``bundle'', if present, would be a transient density fluctuation, not requiring any additional mechanism beyond the local interactions of microtubules (and microtubule turnover), which underlie the basis of the active liquid crystal theory. However, the correlation functions used to quantitatively compare theory and experiments in Figure 3 are all evaluated in the long wavelength limit: i.e. in the low-$q_{x}$ limit to investigate the $q_{y}$ dependence, in the low-$q_{y}$ limit to investigate the $q_{x}$ dependence, and in both the the low-$q_{x}$ and low-$q_{y}$ limits to investigate the $\omega$ dependence. In contrast, many studies of microtubule bundles in spindles analyze the arrangements of microtubules in a cross-section of the spindle \cite{matkovic2022kinetochore, carlini2022coupling, neahring2024torques}.
%(CITE Carlini L, Renda F, Pamula MC, Khodjakov A, Kapoor TM. Coupling of microtubule bundles isolates them from local disruptions to set the structural stability of the anaphase spindle. Proc Natl Acad Sci U S A. 2022 Sep 27;119(39):e2204068119. doi: 10.1073/pnas.2204068119. Epub 2022 Sep 19. PMID: 36122237; PMCID: PMC9522340; Neahring L, Cho NH, He Y, Liu G, Fernandes J, Rux CJ, Nakos K, Subramanian R, Upadhyayula S, Yildiz A, Dumont S. Torques within and outside the human spindle balance twist at anaphase. J Cell Biol. 2024 Sep 2;223(9):e202312046. doi: 10.1083/jcb.202312046. Epub 2024 Jun 13. PMID: 38869473; PMCID: PMC11176257). 
A cross-section is a thin slice, and thus may contain information not present in the long-wavelength limit. We therefore sought to investigate the extent to which the active liquid crystal theory can describe the lateral arrangement of microtubules in spindle cross-sections near the metaphase plate. \\

To test this, we take a constant-$x$ cross-section of the spindle and extract from electron tomography data the $y$-$z$ coordinates $\rvec_{i}$ of each microtubule intersection (Figure.~\ref{Fig4}B). We then evaluate the slice correlation functions,
\begin{equation}
\RCC (\qvec_{\perp} ) = \frac{A}{N_{s}^{2}} \sum_{m, n =1}^{N_{s}} \exp(-i \qvec_\perp  \cdot (\rvec_m - \rvec_n))
\end{equation}
where $A$ is the area of the analysis box, $N_{s}$ is the number of microtubules intersecting the plane and $\qvec_{\perp} = (q_{y}, q_{z})$. The theoretical prediction for this quantity is obtained by marginalizing the 3D spatial correlation function over the wavevector component $q_{x}$ parallel to the spindle-axis (S.I.~3),
\begin{align}\label{eq:Rcc}
& R_{cc}(q_s) = \frac{L_z (\scZero)^2}{4 \sqrt{D} \sqrt{D q_s^2 + \Theta}} + \nonumber\\
&  \frac{L_z (\snZero)^2q_{s''}^2}{8 K'^2}\Bigg [ \frac{
2 \sqrt{2}e_1(e_1^2 -  w^2 )
+ 2q_{s'}^2\left( \alpha -3\sqrt{2}e_1 \right) 
+ \alpha'(q_s)
}{
 D q_{s'}^2 e_1 \left((q_{s'}^2 - e_1^2)^2 - e_1^2 w^2\right)  \alpha(q_s) 
}  \nonumber\\
& + \frac{
2 \sqrt{2}e_2(e_2^2 -  w^2 )
+ 2q_{s'}^2\left( \alpha -3\sqrt{2}e_2 \right) 
+ \alpha'(q_s)
}{
K q_{s'}^2 e_2 \left((q_{s'}^2 - e_2^2)^2 - e_2^2 w^2\right)  \alpha(q_s) 
}  \Bigg ] 
\end{align}
\noindent where $q_s = \vert \mathbf{q}_\perp \vert; K'= D+K; \: R = \sqrt{4q_{s'}^2 + w^2}; \:
\alpha_{\pm} = \sqrt{2q_{s'}^2 + w(w \pm R)};\:
\alpha = \alpha_{+} + \alpha_{-}; \:
\alpha'(q_s) = w \left( R (\alpha_{-} - \alpha_{+}) + w\alpha \right); \:
q_{s'}^2 = q_s^{2} + \frac{\Theta}{D + K}; \:
w^2 = \frac{v_1^2}{(D + K)^2};\:
e_1^2 = q_s^2 + \frac{\Theta}{D}; \:
e_2^2 = q_s^2; \:
q_{s''}^2 = q_s^2 v_1^2.$
% $R = \sqrt{4q_{s'}^2 + w^2}, \alpha_{\pm} = \sqrt{2q_{s'}^2 + w(w \pm R)}, A = A_{+}+A_{-}, \alpha'(q_s) =  w \left( R (\alpha_{-} -  \alpha_{+}) + w\alpha \right), q_{s'}^2 = q_s^{2}+ \frac{\Theta}{D+K}, w^2 = \frac{v_1^2}{(D+K)^2}, e_1^2 =  q_s^2 + \frac{\Theta}{D}, e_2^2 =  q_s^2, q_{s''}^2 = q_s^2 (s_0^n)^2 v_1^2$. 
Using only those fit parameters found previously from spindle geometry or by fitting the long-wavelength correlation functions,  the active liquid crystal theory accurately predicts the cross-sectional density fluctuations at low and intermediate wavenumbers $q_s$ (Fig.~\ref{Fig4}C). However, at the largest wavenumbers, $q_{s} \gtrsim q_{s}^{*} = 20 $ rad$\um^{-1}$, corresponding to real-space distances less than $\lambda^{*} = 2 \pi/q_{s}^* \approx 0.3 \um$, $\RCC(q_{s})$ starts to deviate significantly from the theoretical prediction. As the typical distance between neighboring microtubules in the spindle is $\sim50$ nm \cite{kiewisz2022three}, this corresponds to a few microtubule spacings. The arrangements of microtubules on these length scales are presumably dictated by the molecular properties of cross-linkers in the spindle, and are not captured by the highly coarse-grained active liquid crystal theory. It is an exciting challenge to expand statistical theories of liquids \cite{hansen2013theory} to explain the complex, actively-driven organization of microtubules in the spindle at lengths below $\sim300$ nm. For length scales larger than $\sim300$ nm, which includes those accessible to conventional light microscopy, the active liquid crystal theory is sufficient to quantitatively describe the arrangements of microtubules within the metaphase plate, without the need to invoke specialized bundle-forming mechanisms. \\

% \begin{gather}
%   \label{eq:Rcc}
%     \RCC(q_{r}) =  \frac{L_{z}(\scZero)^{2}}{4\sqrt{D}\,\alpha(q_r)} +\\
% \frac{L_{z}q_r (\snZero)^{2} v_1^2\bigl(\sqrt{D K'}\,q_r+\sqrt{K'}\,\alpha(q_r)+\sqrt{D}\,\beta(q_r)\bigr)}
% {4K(\sqrt{D}q_r+\alpha(q_r))(\sqrt{K'}q_r+\beta(q_r))(\sqrt{K'}\alpha(q_r)+\sqrt{D}\beta(q_r))}, \nonumber
%   \end{gather}
%   where $q_{r} = |\qvec_{\perp}|$, $K'=D+K$, $\alpha(q_{r}) = \sqrt{D q_{r}^{2} + \Theta_{0}}$, $\beta(q_{r}) = \sqrt{K' q_{r}^{2}+\Theta_{0}}$, and $\Theta_{0}' = \Theta_{0}/\sqrt{K'}$. 
\noindent \textbf{Discussion}
This study integrates static micro-mechanical information from high-resolution electron tomography reconstructions with dynamic light microscopy measurements to investigate the physical mechanisms underlying microtubule self-organization in metaphase human tissue culture spindles. Using a coarse-grained active liquid crystal theory, we quantitatively describe the emergent fluctuation spectra of microtubule density and orientation fields. Our findings highlight distinct physical mechanisms driving spindle organization along its long and short axes, providing insights into the anisotropic material properties of the spindle.\\

\indent 
The LC-PolScope has limitations such as projection artifacts, limited spatial resolution, correlated background noise, and low sensitivity to weak birefringence. Similarly, serial-section electron tomography faces challenges like sample preparation difficulties, sectioning artifacts, and alignment issues. Despite this, the fluctuation spectra measured by these two methods show remarkable agreement, highlighting their complementary strengths in studying microtubule organization.\\

\indent 
Our data suggests that nematic elasticity alone governs the relaxation of orientational fluctuations in the spindle along both axes, but that relaxation of microtubule density is more complex. Along the spindle's long axis, microtubule density dynamics are decoupled from orientation and are primarily governed by diffusive-like motion. In contrast, along the short axis, the relaxation of density fluctuations is driven by a combination of alignment interactions (i.e.~nematic elasticity), diffusive-like motion, and microtubule turnover. In particular, the observed deviation of $\scc(q_{0},q_{y})$ from $\propto q_{y}^{-2}$ behavior at high $q_{y}$ suggests that density dynamics are coupled to the director configuration (Fig.~3C, bottom right plot).  Additionally, the non-zero cross-correlation function directly confirms coupling between the fields (Fig.~3C, top right plot).\\

\indent
In our active liquid crystal theory, density-director coupling is captured by the $v_{1} c \Nhat$ term in Eqn.~(\ref{eq:mt_density}), which describes active transport of microtubules along a specific (vector) direction in space, and explicitly breaks the nematic symmetry of the equations of motion. Such symmetry-breaking is anticipated in the LC-PolScope data, as analysis boxes span are centered on one half of the spindle, where coherent polar ordering of microtubules is known to influence transport \cite{mitchison1989polewards, furthauer2019self}. In contrast, the electron tomography data analysis regions are centered at the spindle mid-plane, in which case symmetry in microtubule polarity along the spindle axis is expected. Thus, the observed spindle-scale symmetry-breaking in this region remains unclear. More detailed measurements of the microtubule polarity field (and its fluctuations) may help to answer this question~\cite{brugues2012nucleation, yu2014measuring, bancelin2017probing}. \\

\indent Our work also demonstrates the utility of a phenomenological coarse-grained active liquid crystal framework for inferring material properties, such as nematic elasticity, microtubule diffusivity, and turnover dynamics. The measured parameters provide a quantitative link between microscopic interactions and emergent mesoscale spindle behaviors. Future work should focus on connecting these coarse-grained parameters to molecular-scale interactions to refine the theoretical framework further. Additionally, extending this methodology to study spindle dynamics in other cell types and experimental conditions will provide a broader understanding of the principles governing microtubule organization in diverse biological systems.

\section*{Materials and Methods}
\subsection{Cell lines}
Electron tomography data were obtained from HeLa Kyoto cells, cultured and prepared as described previously~\cite{kiewisz2022three}. For live-cell imaging experiments, we used a stable HeLa cell line expressing EGFP-tagged centrin to fluorescently label centrioles and sfGFP-tagged CENPA to label since centromeres; the establishment and characterization of this cell line was previously reported in \cite{conway2022self}. All cell culture and imaging procedures adhered strictly to protocols detailed in these prior studies.

\subsection{LC-PolScope Imaging, Metaphase Spindle Selection, Definition of Spindle-Referenced Coordinate System}
To acquire live-cell images, we recorded movies consisting of 34 frames captured over approximately 5 minutes. Each movie frame comprised a multiplexed $z$-stack combining LC-PolScope and epifluorescence imaging, which allowed simultaneous visualization of microtubules and centrioles, respectively. To acquire polarization images, LC-PolScope hardware (Cambridge Research Instruments) was mounted on a Nikon TE2000-E microscope equipped with a numerical aperture (NA) 0.52 air condenser lens, a 100×, NA 1.45, oil-immersion objective lens, and a monochromatic bandpass filter centered at $\lambda_{\text{im}} = 546 $ nm. To image centrioles, the microscope was equipped with an LED epifluorescence light source (Thorlabs) and a standard EGFP filter cube. We controlled the microscope hardware using the OpenPolScope MicroManager software package (www.openpolscope.org). \\
Metaphase cells were identified for analysis based on spindle geometry and stability criteria. Specifically, we selected data sets where the interpolar distance remained approximately constant (standard deviation of interpolar distance less than 10\%) throughout imaging. Additionally, to ensure consistent orientation, we included only those data sets in which the vertical positions ($\zhat$-direction) of the centrioles remained within $\pm 1,\mu m$ of each other for the entire duration of the movie. Data sets not meeting these criteria were excluded from further analysis. Live-cell imaging experiments were conducted in four independent experimental replicates, each involving separately plated cells. To correct for spindle translation and rotation in the image plane ($xy$-plane), we performed all analysis in a coordinate system in which the origin is fixed to the center of the spindle and the spindle long axis defines the $\xhat$-direction (Fig.~S3).

\subsection{Calculation of Experimental Correlation Functions}
\paragraph{From Electron Tomography:} To estimate correlation functions from electron tomography reconstructions, we first evaluate the discrete Fourier transforms of the fluctuating director $\Nhat (\Rvec)$ and the normalized density field $C(\Rvec)$ as follows,
\begin{align*}
    \delta C(\mathbf{Q}) &= \frac{V}{\Nseg}\sum_{p=1}^{\NMT} \sum_{m=1}^{M_{p}} \exp(-i \mathbf{Q}\cdot \Rpm ) - \delta (\mathbf{Q}) \\
    \delta \mathbf{N(Q)} &= \frac{V}{\Nseg}\sum_{p=1}^{\NMT} \sum_{m=1}^{M_{p}}\Npm \exp(-i \mathbf{Q}\cdot \Rpm ) - \Nhat_0 \delta (\mathbf{Q})
\end{align*}
where $\mathbf{Q} = (q_x,q_y,q_z)$, and $\Nhat_{0}(\Rvec)$ is the solution to Eqn.~\eqref{eq:NhatSS} with physically realistic boundary conditions, obtained numerically (S.I.~3). We estimate the spatial correlation functions in three dimensions using the direct summation method, i.e.
\begin{align*}
    \SCC(\mathbf{Q}) &= \frac{1}{V}  \delta c\mathbf{(Q)} \delta c\mathbf{(-Q)}  ;\\
    \SNN(\mathbf{Q}) &= \frac{1}{V}   \delta \mathbf{N(Q)} \delta \mathbf{N(-Q)} ;\\
    \SNC(\mathbf{Q}) &= \frac{1}{V}   \delta \mathbf{N(Q)} \delta \mathbf{c(-Q)},
\end{align*}
where $V$ is the volume of the analysis box. In three dimensions, all spatial correlation functions have units of $\sim \um^3$. Two-dimensional, $\zhat$-projected correlation functions are calculated from the 3D ones (Materials \& Methods)
\begin{gather*}
    \scc(\qvec) = \frac{1}{L_{z}}\SCC(q_{x},q_{y},0); \quad
    \snn(\qvec) = \frac{1}{L_{z}}\SNN(q_{x},q_{y},0); \quad \\
    \snc(\qvec) = \frac{1}{L_{z}}\SNC(q_{x},q_{y},0), \quad
  \end{gather*}
  where $L_{z}$ is the average spindle thickness at the metaphase plate (the $x=0$ section of the best-fit ellipsoid), $L_{z} = b^{-1} \int_{-b}^{b}(b^{2}-y^{2})\text{d}y = \pi b/2$. The two-dimensional, $\zhat$-projected correlation functions have units of $\sim \um^2$.\\

  \paragraph{From LC-PolScope measurements:}
  To estimate correlation functions from LC-PolScope data, we first compute the normalized density field $c(\rvec,t) = r(\rvec,t)/ \langle r \rangle_{t}(\rvec)$. The time-averaged retardance field $\langle r \rangle_{t}(\rvec)$ is calculated by fitting the computed time-averaged retardance image to (Fig.~\ref{Fig2}B, middle row) to a paraboloid. We use Mathematica’s inbuilt Fourier[] function to approximate the (2 + 1)D Fourier transforms of the fluctuations $\delta c = c(\rvec,t) - 1$, $\dcT(\qvec, \w)$ and $\dnyT(\qvec, \w)$. To mitigate boundary effects, we apply mirror padding in space and time to all data arrays before computing all self-correlation functions (nn, $cc$ correlations) \cite{gonzales2018digital}.  We omit the mirror padding step when computing the cross-correlations (n$c$ correlations), since mirror padding forces the imaginary component of the cross-correlation to vanish artificially. After computing the Fourier transforms $\dcT(\qvec, \w)$ and $\dnyT(\qvec, \w)$, we calculate the spatiotemporal and equal-time correlation functions:
\begin{gather*}
  \ccc(\qvec, \w) = \frac{1}{\tau_{0}\lambda_{0}^{2}} \dcT(\qvec, \w) \dcT(-\qvec, -\w);\\
  \cnn(\qvec, \w) = \frac{1}{\tau_{0}\lambda_{0}^{2}} \dnyT(\qvec, \w) \dnyT(-\qvec, -\w);\\
  \cnc(\qvec, \w) = \frac{1}{\tau_{0}\lambda_{0}^{2}} \dnyT(\qvec, \w) \dcT(-\qvec, -\w);\\
  \scc(\qvec) = \frac{1}{2\pi}\int \ccc(\qvec, \w) \dd \w; \quad  \snn(\qvec) = \frac{1}{2\pi}\int \cnn(\qvec, \w) \dd \w;\\
  \snc(\qvec) = \frac{1}{2\pi}\int \cnc(\qvec, \w) \dd \w.
\end{gather*}
where $\tau_{0}$ is the total time of the movie being analyzed and the $\w$-integrals are taken over all available angular frequencies $\w$. To correct for Gaussian-distributed background noise in the retardance and slow-axis signals, we initially fit each temporal correlation function at fixed wavevector to functions of the form:
\begin{gather*}
  \cnn(\qvec_{0}, \w) = \frac{A_{\text{nn}}}{\w^{2}} + B_{\text{nn}};\\
  \ccc(\qvec_{0}, \w) = \frac{A_{cc}}{\w^{2}} + B_{cc}.
\end{gather*}
We then subtract the fitted additive constants from the original correlation function,
\begin{gather*}
  \cnn(\qvec_{0}, \w)  \rightarrow \cnn(\qvec_{0}, \w)  -  B_{\text{nn}};\\
  \ccc(\qvec_{0}, \w) \rightarrow \ccc(\qvec_{0}, \w) - B_{cc}.
\end{gather*}

To correct for spurious high-$q$ correlations induced by the finite optical resolution of the LC-PolScope, we deconvolve the correlation functions $\scc(\qvec)$, $\snn(\qvec)$, and $\snc(\qvec)$ by the experimentally measured optical impulse function of the LC-PolScope (Fig.~S6), 
\begin{gather*}
\scc(\qvec) \rightarrow e^{\sigma_{0}^2\qvec^{2}}\scc(\qvec); \qquad
\snn(\qvec)  \rightarrow e^{\sigma_{0}^2\qvec^{2}}\snn(\qvec);\\
\snc(\qvec)  \rightarrow e^{\sigma_{0}^2\qvec^{2}}\snc(\qvec),
\end{gather*}
where $\sigma_{0} = 61 $ nm characterizes the point spread function. Finally, we discard data for
which $|\qvec| > 2 \pi/\lambda_{\text{im}}$, where $\lambda_{\text{im}} = 265$ nm is half the
microscope illumination wavelength.

\subsection{Model Fitting \& Parameter Estimation}
The predictions of the model (Eqns.~\eqref{eq:FourierCFs}) were fitted to the log-transformed experimental data (black points in Fig.~3). Best-fit values were obtained by minimizing the sum of squared residuals across all log-transformed data sets simultaneously. Parameter uncertainties were estimated via non-replacement subsampling~\cite{bickel2011resampling}: in each trial, we randomly selected 6 of the 11 LC-PolScope single-spindle data sets and 2 of the 3 single-spindle ET data sets. For the spatial correlation functions, we averaged the selected data sets, giving equal weight to LC-PolScope and ET data; the model was then fitted to the composite subsampled data set. This procedure was repeated 50 times, and the means and standard deviations of the distributions of best-fit parameters were estimated. For each parameter, the reported uncertainty (Main Text and Table 1) is the larger of the following two quantities: (i) the standard deviation over subsamples; (ii) the absolute difference between the mean of the subsample distribution, and the best-fit value obtained by fitting all data sets simultaneously. 

To estimate the uncertainty in $\RCC(q_{s})$ at each wavenumber $q_{s}$ (Fig.~4C, black dashed curves), we recomputed the function after setting every fit parameter to the upper and lower bounds of its estimated range (Table 1). Since there are five fit parameters ($\scZero$, $\snZero$,  $D$, $K$, and $v_{1}$), this generated $2^5 = 32$ realizations of $\RCC(q_s)$ for each $q_s$. The standard deviation of these 32 values was taken as the uncertainty for that wavenumber.

\subsection{Data \& Code Availability}
All data and codes used in this paper will be made available on request.% \cite{HeLaZenodoDatabase}.\\

\section*{Acknowledgments}
We acknowledge support from the CCB$_X$ program of the Center for Computational Biology of the 
Flatiron Institute. We thank Robert Kiewisz for providing his (published) data on the 3D organization of MTs in the HeLa cells. We thank Michael O'Brien (Harvard University) for help preparing the \textit{in vitro} microtubule bundles, and Will Conway (New York Center for Structural Biology) for sharing unpublished microtubule polymerization data in HeLa spindles. Research in the M\"{u}ller-Reichert lab is supported by the Deutsche Forschungsgemeinschaft (DFG, grant number 258577783) and the CCB$_X$ program (SF, 1157392).

\onecolumngrid
\clearpage

\clearpage

\clearpage
\begin{figure*}[t!]
  \includegraphics[width=16cm]{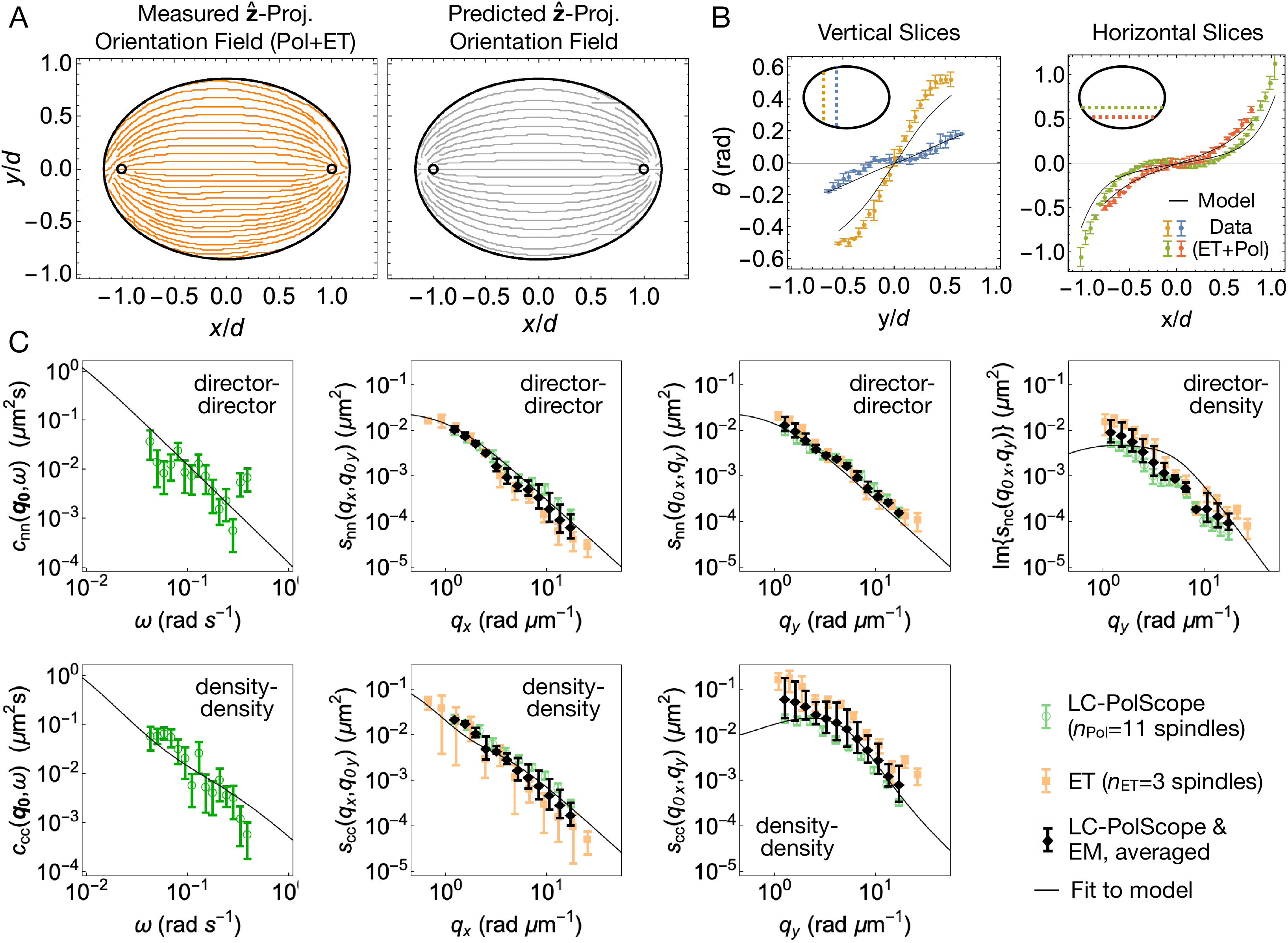}
  \caption{Active liquid crystal theory theory accurately describes microtubule orientation in HeLa spindles, as well as spatiotemporal correlations in microtubule density and orientation. (A) \textit{Left:} The $\zhat$-projected orientation field and best-fit spindle boundary, averaged over both electron tomography (ET) and LC-PolScope data sets. Prior to averaging, $x$- and $y$-coordinates of all data sets were normalized by the interpolar half-spacing $d$. \textit{Right:} Numerical prediction of $\zhat$-projected orientation field, found by solving Eqn.~\ref{eq:NhatSS}, supplemented with boundary conditions based on the average spindle geometry (S.I.~3). (B) Vertical (\textit{left}) and horizontal (\textit{right}) slices through the data shown in A. Error bars represent the discrepancy between LC-PolScope and electron tomography estimates; solid black lines show theoretical predictions. (C) Selected director-director, density-density, and director-density correlation functions, for LC-PolScope data (green circles), electron tomography data (ET, yellow squares), and averaged (black diamonds). Error bars for LC-PolScope and electron tomography data are given by the standard error over individual-spindle data sets; error bars in the averaged correlation functions indicate the discrepancy between the two measurements. For the spatiotemporal correlations $\cnn(\qvec_{0}, \w)$ and $\ccc(\qvec_{0}, \w)$ (\textit{left column}), only LC-PolScope data is available. The thin black line shows fits to Eqns.~\eqref{eq:FourierCFs}, with parameters reported in Table 1.}
\label{Fig3}
\end{figure*}

\clearpage
\begin{figure*}[t!]
  \includegraphics[width=16cm]{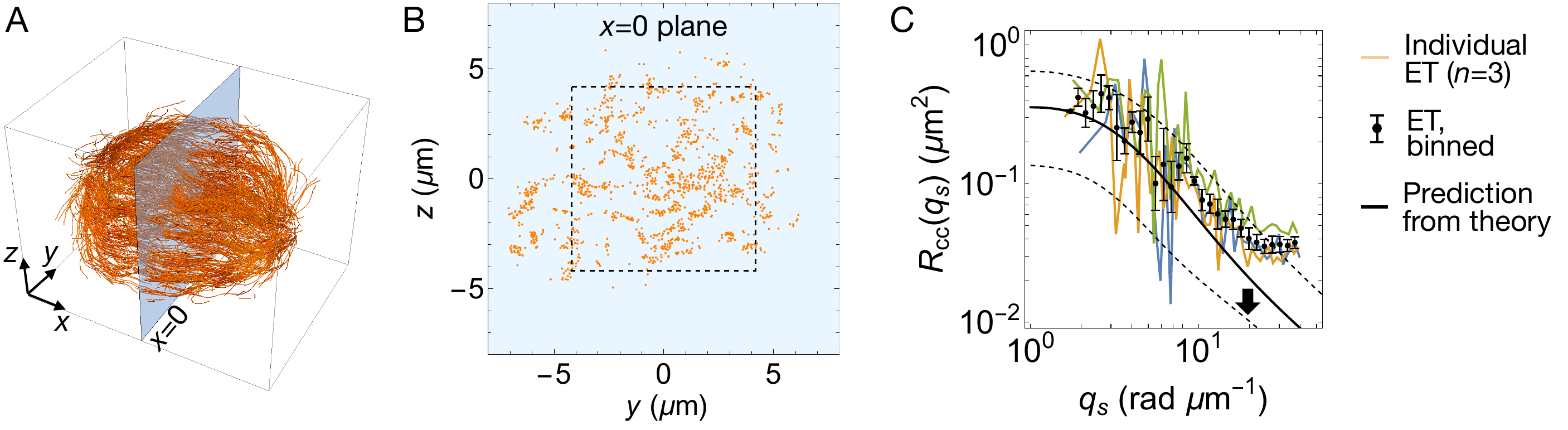}
  \caption{Arrangement of microtubules in spindle cross-sections are predicted by the active liquid crystal theory without additional fitting parameters. (A) Correlations in microtubule density are analyzed in the slice $x=0$ perpendicular to the spindle axis, obtained from electron tomography reconstructions. (B) Each orange dot indicates the intersection of a microtubule with the plane $x=0$; correlation analysis is restricted to the largest square region fully contained within the spindle boundary (indicated by dashed black lines). (C) Experimentally determined radial correlations $\Rcc(q_{s})$ from electron tomography data (ET, black points with error bars) are well fitted by the predictions of the active liquid crystal theory (solid black curve), using only fit parameters previously obtained from correlation functions in the long-wavelength limit (Table 1). Black dashed curves indicate the range of uncertainty in $R_{CC}(q_{s})$ consistent with estimated uncertainties in $\scZero$, $\snZero$,  $D$, $K$, and $v_{1}$ (\textcolor{blue}{Materials \& Methods}). The prediction agrees well with the data for wavenumbers $q_{s} \lesssim 20$ rad $\um^{-1}$ (black arrow), corresponding to distances greater than $\lambda^* = 2 \pi/q_{s}^{*} \approx 0.3 \um$. Error bars indicate the standard error of the data in each bin.}
\label{Fig4}
\end{figure*}

\clearpage
\setlength{\tabcolsep}{10pt} % Default value: 6pt
\renewcommand{\arraystretch}{1.5}
\begin{table}
\begin{tabular}{ |c|c|c|c| }
 \hline
  Parameter Name &  Symbol & Value & Units \\
  \hline \hline
  Concentration noise amplitude & $\scZero$ & $0.0265 \pm 0.0039$ & $\um \,s^{-1/2}$ \\
  \hline
Director noise amplitude & $\snZero$ & $0.0110 \pm 0.0005 $ & $\um \, s^{-1/2}$ \\ 
 \hline
 Microtubule diffusion constant & $D$ & $0.0043 \pm 0.0023$& $\um^2s^{-1}$ \\ 
 \hline
  Nematic diffusivity & $K$ & $0.0021 \pm 0.0002 $& $\um^2s^{-1}$\\
   \hline
 Polar microtubule transport speed &$v_{1}$ & $ 8.4 \pm 6.2$  & $\um$ min$^{-1}$ \\ 
 \hline
Microtubule turnover rate & $\Theta_{0}$ & 0.047 & $s^{-1}$ \\ 
  \hline
\end{tabular}
\caption{Table of best-fit parameters for spindle material properties. The value of $\Theta_{0}$ was calculated from a weighted average of previously-measured kinetochore and non-kinetochore microtubule catastrophe rates~\cite{conway2022self}. Estimates are found by fitting combined electron tomography and LC-PolScope correlation functions (Fig.~\ref{Fig3}); uncertainties are found via non-replacement subsampling \textcolor{blue}{Materials \& Methods}.}
\label{tab:parameters}
\end{table}

\clearpage

%% file: src/supplementary.tex
\section{Electron Tomography Data}
\subsection{Identification of Poles \& Definition of Spindle Coordinate System}
We identify the spindle poles with the positions of the centrioles, which are obtained directly from electron tomography data~\cite{kiewisz2022three}. For each data set, the straight line joining the poles defines the spindle long axis ($x-$axis) and the distance between centrioles is taken as the interpolar distance $2d$. The $y-$ and $z-$ axes are chosen arbitrarily.

\subsection{Description of Microtubule Data}
A complete electron tomography reconstruction consists of $\NMT$ microtubules. Microtubule $i$ is described by an ordered set $\Pim$, where  $1 \leq m \leq M_{i}+1$ points; the order of the points indicates their distance from one reference end of the microtubule (chosen arbitrarily), measured along tubule arc-length.  All subsequent calculations are performed in a coordinate system where the origin is placed at the center of mass of the microtubule point cloud.  The centroid $\Rim$ and orientation $\Nim$ of each the sub-segments in microtubule $i$ is given by:
\begin{equation}
   \Rim = \frac{1}{2} \left( \Pim + \PimA \right)  \qquad  \Nim = \frac{\PimA - \Pim}{\left| \PimA - \Pim \right|}.
 \end{equation}
Some basic statistics of the three electron tomography data sets are given in Fig.~\ref{sifig:density-profiles}C; more detailed analysis, including details of microtubule length distribution, curvature, and properties of kinetochore fibers, is given in references \cite{conway2022self, kiewisz2022three}.

\subsection{Rescaling Spindle Coordinates to Match LC-PolScope Estimates of Pole Position}
\label{et-rescaling}
We observed that the average interpolar half-distance $d$ calculated from our three electron tomography data sets is smaller than the same quantity estimated from LC-PolScope data. We also noticed that our initial estimates of spatial correlation functions, computed from electron tomography data, were correspondingly displaced along the horizontal axis. These discrepancies could reflect random sample-to-sample variation, or could be due to systematic artifacts associated with electron tomography sample preparation or data acquisition. Freeze-substitution electron microscopy is known to cause significant sample shrinkage and deformation~\cite{luther1988method}. While we use Z-corrected ET coordinates correct for this artifact~ \cite{kiewisz2022three} , the applied correction may not be fully accounted for sample shrinkage and deformation. To account for the possibility of systematic error in the ET scaling, we defined a rescaling factor $\psi$,
\begin{equation}
  \psi = \frac{\langle d \rangle_{\text{Pol}}}{\langle d \rangle_{\text{ET}}} \approx  1.44.
\end{equation}
All subsequent analyses of ET data uses rescaled coordinates, $\Rim \rightarrow \psi \Rim$.

\subsection{Fitting Spindle Boundary to Prolate Ellipsoid}
\label{sisec:et-boundary}
To estimate the dimensions of a single spindle from electron tomography data, we fit a prolate ellipsoid (with long and short semi-axes $a$ and $b$ respectively) to the spindle boundary by finding the minimum-volume axis-aligned ellipsoid that contains 90\% of all points $\Rim$ for $1 \leq m\leq M_{i}$ and $1 \leq i \leq \Nseg$.

\subsection{Microtubule Cross-Sectional Density is Constant to Within $\sim 20\%$ in the Analysis Volume}
To investigate the extent to which microtubule density varies within the spindle, we plot the cross-sectional density $\rho_{2D}$ as a function of position along the spindle long axis (Fig.~\ref{sifig:density-profiles}). To estimate $\rho_{2D}(x)$, we compute the number of microtubules that intersect in the plane of constant $x$, and divide by the area of the disk defined by the intersection of that plane with the best-fit ellipsoid. All data sets show dips of around $\sim 20\%$ near $x=0$; this presumably corresponds to the presence of metaphase chromosomes, which occupy a significant amount of the spindle volume and thus reduce the average microtubule density at the metaphase plate (Fig.~\ref{sifig:density-profiles}A).

\subsection{$\zhat$-Projected Microtubule Orientation in Individual Electron Tomography Data Sets is Consistent with Predictions of the Active Liquid Crystal Theory}
\begin{figure}[t!]
  \includegraphics[width=\textwidth]{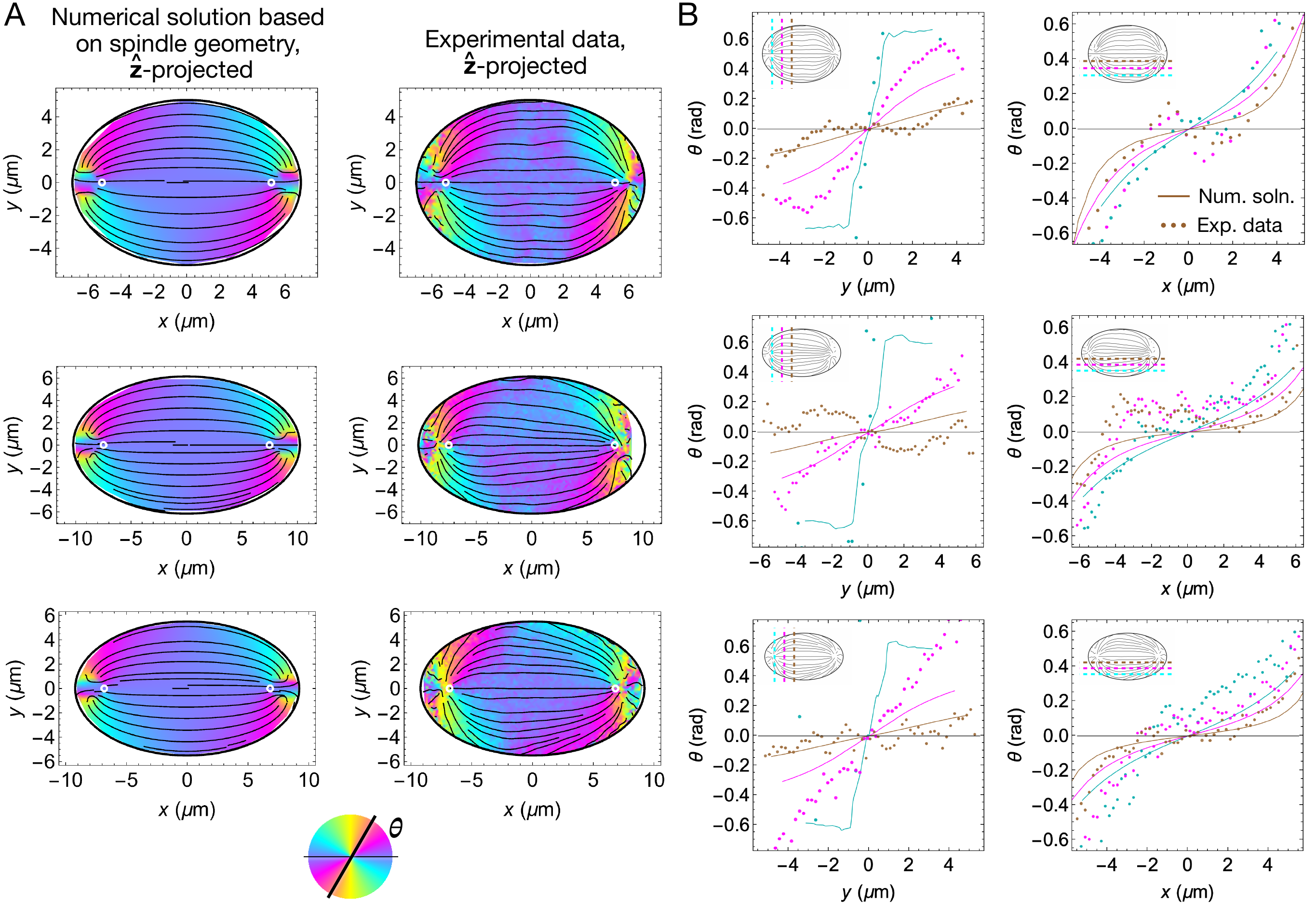}
  \caption{Comparison of microtubule orientation, measured by electron tomography, with predictions of the active liquid crystal model. (A) Numerical solution of Main Text Eqn.~\ref{eq:NhatSS} (left) with $\zhat$-projected orientation fields (right) for each of the three electron tomography data sets we analyzed. In all images, white circles indicate the positions of the centrioles. In the images on the right, data is only shown inside the ellipse that defines the spindle boundary (S.I.~\ref{sisec:et-boundary}). In all images, thin lines in the interior of the ellipse show streamlines of the projected orientation field. (B) Vertical and horizontal slices of the experimentally measured projected orientation field (points) show good agreement with predictions of the numerical solution (thin lines). Vertical slices are taken at $y = -\frac{3a}{4}, -\frac{a}{2}$; horizontal slices are taken at $-\frac{a}{4}$ and $x = -\frac{3b}{4}, -\frac{b}{2}$, and $-\frac{b}{4}$.}
\label{sifig:et-vs-model}
\end{figure}

\begin{figure}[t!]
  \includegraphics[width=\textwidth]{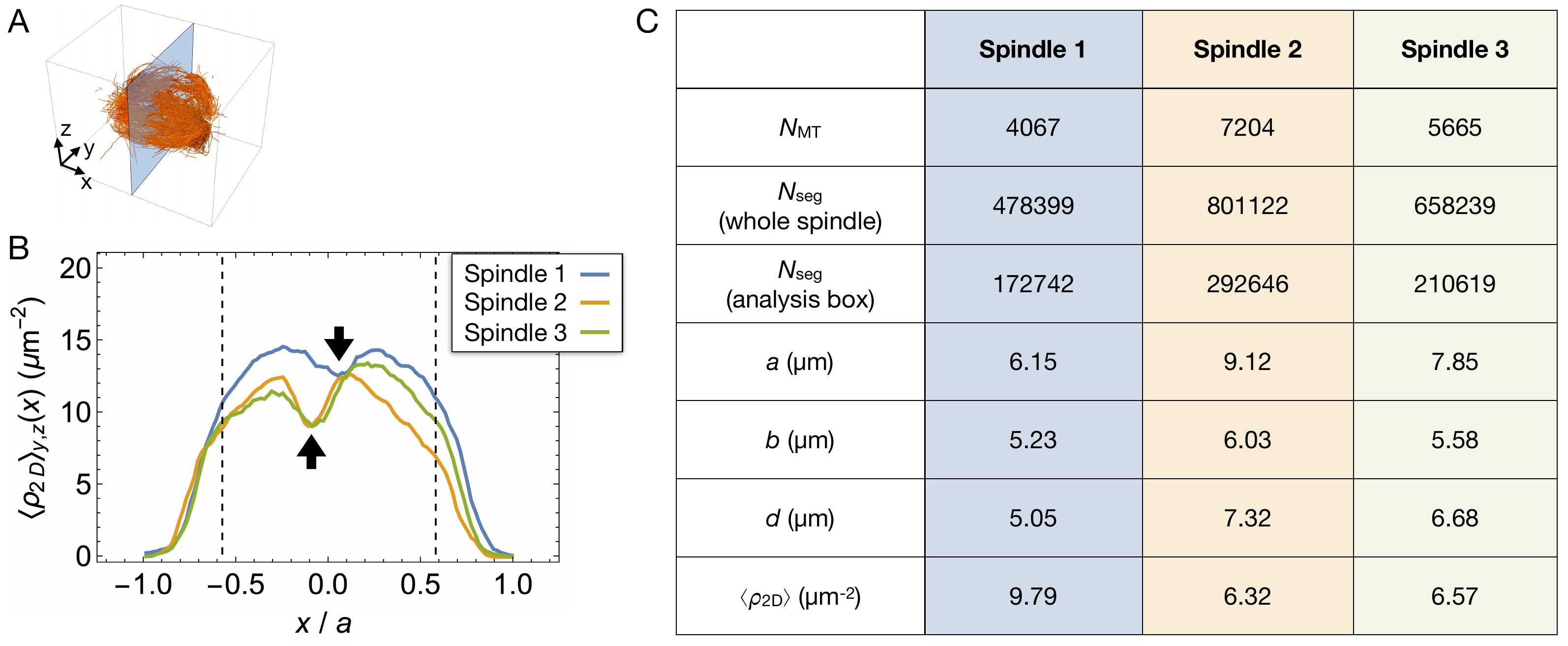}
  \caption{Cross-sectional microtubule density profiles and summary statistics for the three electron tomography-reconstructed spindles we analyzed. (A) The microtubule cross-section density $\langle \rho_{2D} \rangle_{y,z} (x)$ is computed by calculating the number of microtubules that intersect with a constant-$x$ plane, and dividing by the cross-sectional area of the spindle at that position. (B) Plots of 
 $\langle \rho_{2D} \rangle_{y,z} (x)$ for each of the three electron tomography reconstructions; $x$-coordinates are plotted relative to each spindle's long semi-axis $a$. The dip near the origin corresponds to the presence of chromosomes at the metaphase plate. Black dashed lines indicate the boundaries of the analysis box, which has dimensions $\frac{2a}{\sqrt{3}} \times \frac{2b}{\sqrt{3}} \times \frac{2b}{\sqrt{3}} $. (C) Shape and microtubule density statistics for each spindle. The average density $\langle \rho_{2D} \rangle$ is computed in the analysis box, i.e. between the dashed vertical lines in panel B.}
\label{sifig:density-profiles}
\end{figure}

It was previously shown that the microtubule orientation in HeLa spindles, as estimated from electron tomography data, is well-described by the predictions of the active nematic model \cite{conway2022self}. To independently verify this result and to facilitate direct comparisons with LC-PolScope data (S.I.~\ref{sisec:polscope}), we project the measured directors $\Nim$ in each electron tomography data set over $\zhat$ and re-normalized to obtain a discrete 2D director field $\nim$,
\begin{equation*}
\nim = \frac{\mathcal{P}_{z}\{ \Nim \}}{|\mathcal{P}_{z}\{ \Nim \}|}.
\end{equation*}
To remove the dependence of our estimate on the arbitrarily chosen projection axis ($\zhat$-axis), we perform a series of 10 rotations of the data about $\xhat$, equally spaced between 0 and $2\pi$, of to obtain 10 different estimates of $\nim$ and average over all resulting projections. Projected orientation fields are well-fitted by the results of numerical solutions based on the geometry of individual spindles (Sec.~\ref{sisec:numerical-solution}, Fig.~\ref{sifig:et-vs-model}).

\section{LC-PolScope Data}
\label{sisec:polscope}
\subsection{Definition of Spindle Coordinate System}
\begin{figure}[t!]
\includegraphics[width=0.7\textwidth]{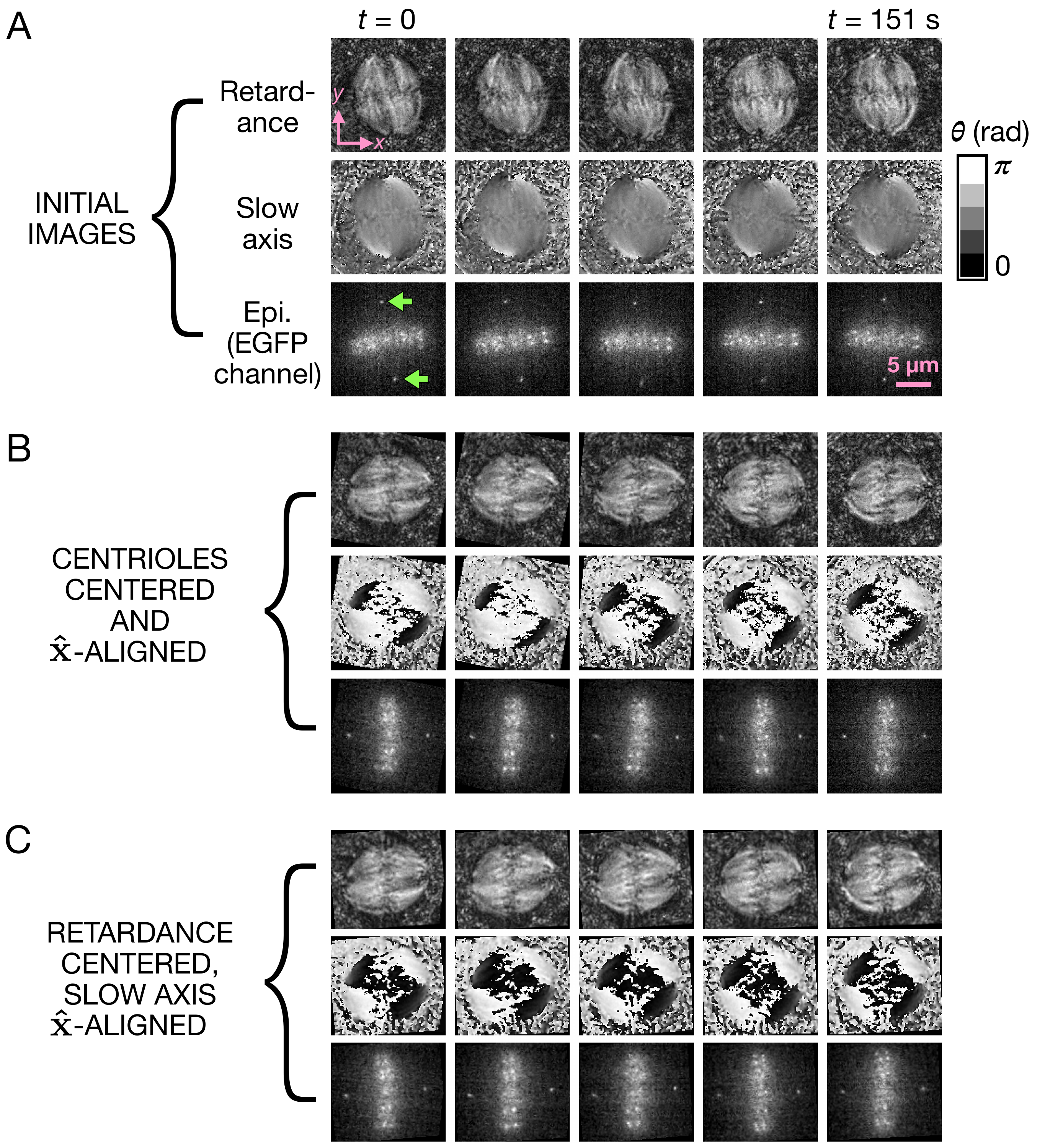}
\caption{Definition of spindle-referenced coordinate system for LC-PolScope images. (A) Original images from a sample time series, showing retardance, slow axis, and epifluorescence. Green arrows in the $t=0$ epifluorescence image indicate the centrioles (EGFP:centrin). (B) Results of the first stage of coordinate system identification, where the average position of the centrioles is positioned at the origin (center of image), and the line joining the centrioles lies along the $\xhat$-axis. (C) Results of the later stages of image alignment, in which the center of brightness of the spindle (i.e. the centroid of a binarized mask of the retardance image) defines the origin, and the spindle is rotated such that the average angle inside the body of the spindle defines the $\xhat$-axis.
}
\label{sifig:polscope-coords}
\end{figure}

Our algorithm to define a spindle-referenced coordinate system for the LC-PolScope data comprises several steps (Fig.~\ref{sifig:polscope-coords}). First, we perform an initial transformation of the LC-PolScope images such that, in each frame, the origin is defined as the average position of the two centrioles (measured via epifluorescence), and the $\xhat$-axis is defined to be parallel to the line joining them. Due to finite time lag of the LC-PolScope hardware, several seconds go by between acquisition of the epifluorescence image and acquisition of the the associated retardance and slow axis images. Consequently, we find that our preliminary calculation of the spindle coordinate system does not adequately center or align these channels (Fig.~\ref{sifig:polscope-coords}B). We therefore use the LC-PolScope retardance and slow-axis images to re-center and re-orient the spindle. To do this, we manually crop a region of interest (ROI) that includes all positions of the spindle over the course of the entire movie. Next, we apply a Gaussian blur (radius 2 $\um$) to each frame of the retardance movie and binarize the resulting images using a threshold of half the maximum intensity in the cropped image. Images are translated such that the center of brightness of the largest contiguous white region is at the origin. Finally, we use the slow axis data to rotate the spindle such that its average orientation coincides with the horizontal axis of the image (Fig.~\ref{sifig:polscope-coords}C). We use this spindle-referenced coordinate system for all subsequent calculations.

\subsection{Fitting Spindle Boundary to Ellipse}
\label{sisec:polscope-boundary}
\begin{figure}[t!]
\includegraphics[width=\textwidth]{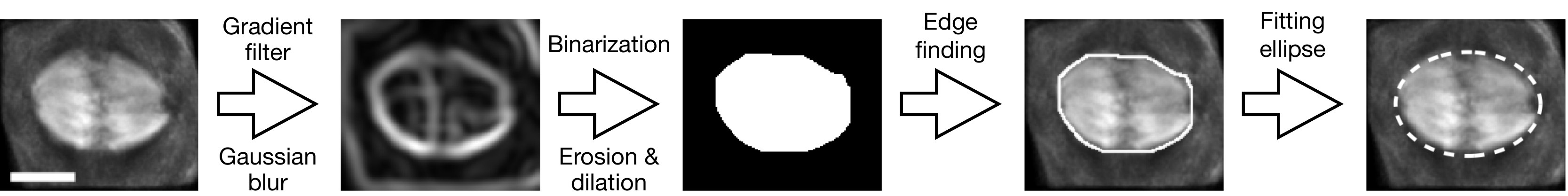}
\caption{Fitting the spindle boundary from LC-PolScope retardance image. From left to right: we begin with the time averaged retardance image, apply a gradient filter (radius $1 \um$) followed by a Gaussian filter (radius $0.5 \um$). We then apply a local adaptive binarization algorithm (radius $4 \um$), followed by erosion ($1.5 \um$), dilation ($1 \um$), and a filling transform. We then find the edge of the white region (solid white curve), and fit it to an axis-aligned ellipse (dashed white curve). Scale bar $5 \um$.
}
\label{sifig:polscope-boundary}
\end{figure}

For each LC-PolScope movie, we fit the boundary of the spindle to an ellipse with long and short semi-axes $a$ and $b$ respectively. To robustly find the spindle boundary, we apply a series of classical image processing functions -- Gaussian convolution, gradient filtering, erosion, dilation, and adaptive binarization -- to the LC-PolScope retardance image. These algorithms are implemented in the Mathematica programming language; we use the same set of image processing parameters in our analysis of all data sets. The complete edge-identification algorithm, together will all image processing parameters used, is shown in Fig.~\ref{sifig:polscope-boundary}. Finally, we fit the boundary points to an axis-aligned ellipse, defined by
\begin{equation*}
  \Big(\frac{x}{a} \Big)^{2}+  \Big(\frac{y}{b} \Big)^{2} = 1.
  \end{equation*}

\subsection{Comparison of Spindle Shape from LC-PolScope Retardance Images with Spindle Shape from Electron Tomography Data}
Table \ref{sitab:paramcomparison} shows a comparison of the average best-fit ellipse parameters found by LC-PolScope with the best-fit ellipsoid parameters obtained from electron tomography data (S.I.~\ref{sisec:et-boundary}). The LC-PolScope estimate for the spindle long semi-axis $a$ is $\sim 25 \%$ smaller than the corresponding electron tomography value, while the LC-PolScope estimate of the spindle short axis is $\sim 25 \%$ larger. This difference is presumably due to the lower sensitivity of the LC-PolScope to microtubules near the spindle poles compared to microtubules near the central spindle; the former tend to be less perpendicular to the optical axis $\zhat$ and thus contribute less to the retardance signal (next section).

\subsection{Time-Averaged Retardance and Orientation Fields from LC-PolScope Data Are Consistent with Predictions of the Active Liquid Crystal Theory}
\label{sisec:polscope-vs-model}
\begin{figure}[t!]
\includegraphics[width=16cm]{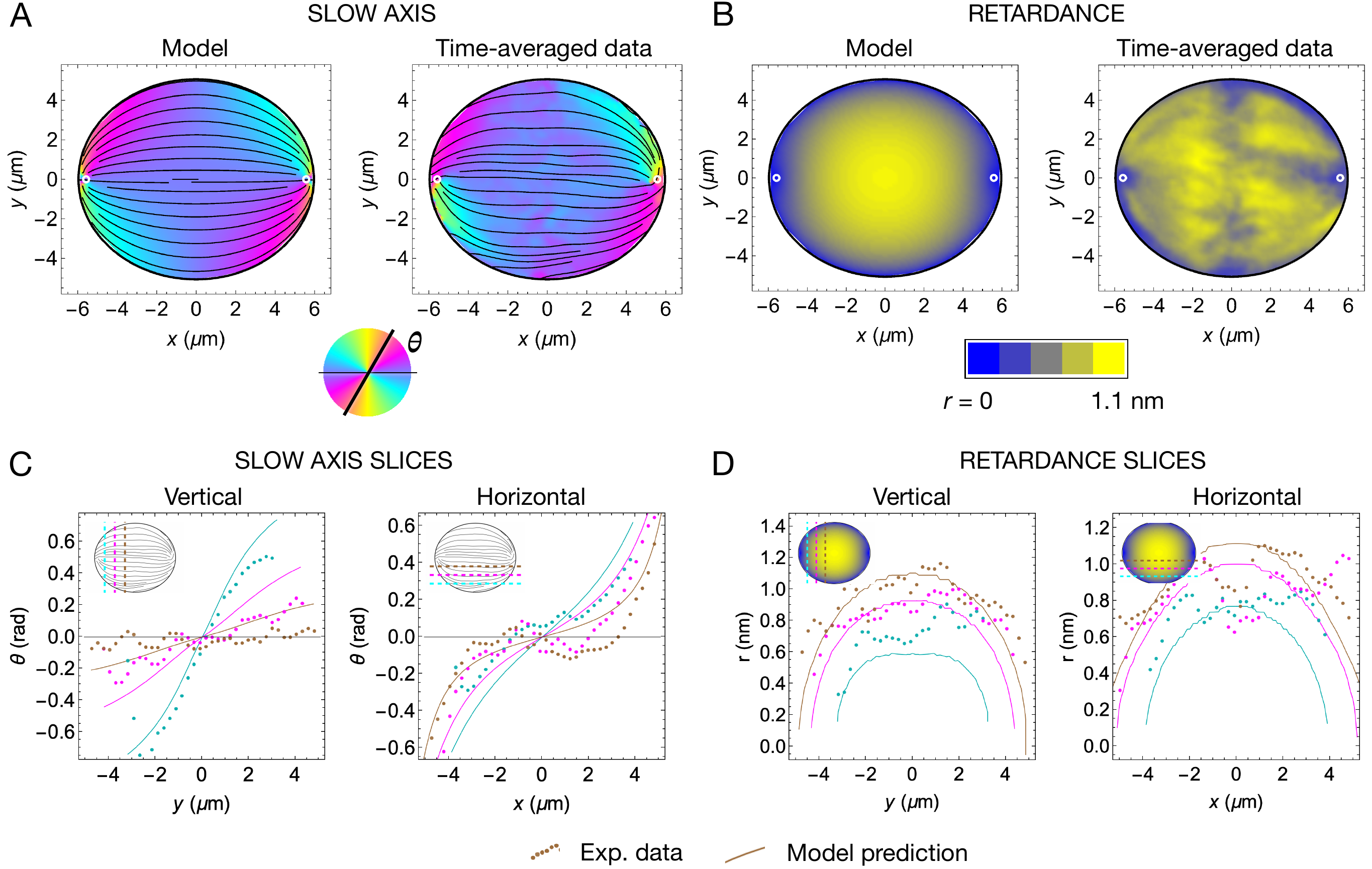}
\caption{Comparison of LC-PolScope data with best-fit numerical solutions of the model for a singe spindle. (A) Time-averaged LC-PolScope orientation data for a typical spindle, plotted alongside the prediction of the numerical solution of the model (Eqn.~\ref{eq:NhatSS}), using the geometry of that spindle to define boundary conditions. In the numerical solution, the defect position [white circles at $(\pm 5.6 \um, 0)$] is treated as a fit parameter; it is chosen to minimize the difference between the experimentally observed and the numerically predicted orientation field. (B) Time-averaged LC-PolScope retardance data for the same spindle, plotted alongside the predictions of the numerical solution. To scale the numerical solution model to the data, we use a single fit parameter equivalent to the average microtubule cross-sectional density $\langle \rho_{2D} \rangle$ (S.I.~\ref{sisec:polscope-vs-model}). (C \& D) Vertical and horizontal slices through the slow axis and retardance data. Solid lines show the predictions of the model, points show LC-PolScope data. Insets in the top left of each plot indicate the positions of the slices. Vertical slices are taken at $x = -\frac{3a}{4}$, $-\frac{a}{2}$, and $-\frac{a}{4}$; horizontal slices are taken at $y = -\frac{3b}{4}$, $-\frac{b}{2}$, and $-\frac{b}{4}$, where $a$ and $b$ are the long and short semi-axes of the ellipse that best fits the spindle boundary (S.I.~\ref{sisec:polscope-boundary}). 
}
\label{sifig:polscope-vs-model}
\end{figure}

It was previously shown that the microtubule orientation in HeLa spindles, as estimated from LC-PolScope data, is well-described by the predictions of Eqn.~\ref{eq:NhatSS}~\cite{conway2022self}. To independently replicate this finding, we performed a series of numerical solutions of Eqn.~\ref{eq:NhatSS} (S.I.~\ref{sisec:numerical-solution}). For each data set, we supplemented Eqn.~\ref{eq:NhatSS} with a boundary condition that forces the director to lie tangent to the 3D prolate ellipsoid with long and short semi-axes $a$ and $b$, obtained as described in the previous section. In these simulations, we treated the interpolar half-distance $d$ as a fit parameter, selecting the best fit spindle by minimization of the quantity
\begin{equation*}
  \int_{A}|\exp{(2i \theta(x,y))} - \exp{(2i \theta_{0}(x,y))} |^{2} \dd A,
\end{equation*}
$\theta_{0}(x,y)$ is the simulated angle of the 2D nematic field found by projecting the 3D solution over the $\zhat$-axis, and the integral is evaluated over the area $A$ inside the spindle boundary. Using the best-fit value of $d$, the predicted slow axis field $\theta_{0}(x,y)$ provides good fits to time-averaged slow axis measurements from individual spindles (Fig.~\ref{sifig:polscope-vs-model}A\&C).

To compare the predictions of the simulation with our retardance data, we compute the expected retardance profile using an and-hoc generalization of Eqn.~\ref{eq:retard} that takes into account the fact that microtubules perpendicular to the optical axis ($\Nhat \cdot \zhat \approx 0$) contribute maximally to the retardance signal, while microtubules parallel or antiparallel to it ($\Nhat \cdot \zhat \approx \pm 1$) do not contribute,
\begin{equation*}
   r(\rvec, t) \approx A_{0} \int_T \cos(2 \phi) \rho_{2D}(\Rvec, t) d z; \label{eq:retardFull},
   \end{equation*}
   where $\phi = \arccos{(\Nhat \cdot \zhat)}$. To facilitate comparison of this relation with LC-PolScope data, we make the further approximation $\rho_{2D}(\Rvec, t) = \langle \rho_{2D} \rangle$, which we take to be constant in space and time. Since $\phi$ is fully determined by the numerical simulations (using the previously-fitted value of $d$), and $A_{0}$ is a known constant, the only remaining fit parameter is $\langle \rho_{2D} \rangle$. The retardance signal shows significant departures from the predicted form (Fig.~\ref{sifig:polscope-vs-model}B\&D). This is presumably due to structural features, such as metaphase chromosomes, that cause persistent deviations from uniform density. Nevertheless, important large-scale patterns, such as higher retardance near the central spindle than at the boundaries, are captured by the model. We can also compare the fitted values of  $\langle \rho_{2D} \rangle$, averaged over all spindles we imaged, to those directly measured from electron tomography data (Fig.~\ref{sifig:density-profiles}). These values agree within experimental uncertainty: $\langle \rho_{2D} \rangle = (13.6 \pm 0.7) \um^{-2}$ from LC-PolScope retardance data, versus  $\langle \rho_{2D} \rangle = (11.7 \pm 0.8) \um^{-2}$ from ET data (mean $\pm$ standard error, calculated over $n_{\text{Pol}}=11$ spindles for LC-PolScope and $n_{\text{ET}}=3$ spindles for electron tomography). Student's T-Test (two-tailed) gives $p = 0.19$, indicating no statistically significant difference between the means of the ET and LC-PolScope data sets.
 
\subsection{Estimating the Lateral Optical Impulse Function of LC-PolScope}
At length scales comparable to the illumination wavelength, $\lambda_{\text{im}} = 546$ nm, spurious correlations are induced in LC-PolScope data due to diffraction. To correct for these artifacts, we deconvolved our calculated spatial correlation functions with the lateral optical impulse function (OIF) of the LC-PolScope; this quantity is equivalent to the lateral point spread function (PSF) in fluorescence microscopy. Unfortunately, no rigorous theory exists that allows straightforward estimation of the OIF for ``semi-coherent'' imaging techniques like LC-PolScope~\cite{tran2022point}. We therefore estimated the lateral OIF for our system experimentally by imaging \textit{in vitro}-assembled microtubule bundles with diameters much less than $\lambda_{\text{im}}$. To manufacture these bundles, we first followed previously established protocols to prepare a suspension of long (average length $\gtrsim 2 \um$), stabilized microtubules~\cite{tayar2022assembling}. Briefly, we prepared a solution containing 0.6 mM GMPCPP, 1.2 mM dithiothreitol (DTT), and 0.2 mg/mL Alexa-647-labeled tubulin in M2B buffer. After preparing the mixture as described in~\cite{tayar2022assembling}, we incubated it for an additional 8-12 hours at room temperature. To promote the formation of microtubule bundles, we next diluted the stabilized microtubule suspension $10 \times$ with room temperature M2B buffer, and immediately added polyethylene glycol 20 kD to a final concentration of $1\%$ w/v. After incubating for at least 1 hr, we confined $2$-$5 \mu$L of the suspension between parallel cover slips; this results in cover slips spaced $\sim 10 \um$ apart. As the droplet spreads between the cover slips, some microtubule bundles spontaneously and irreversibly stick to the enclosing glass surfaces. The suspension-filled chamber is sealed with UV-cure glue (Norland) and placed on a microscope equipped with LC-PolScope and epifluorescence modes.

\begin{figure}[t!]
  \centering
  \includegraphics[width=\textwidth]{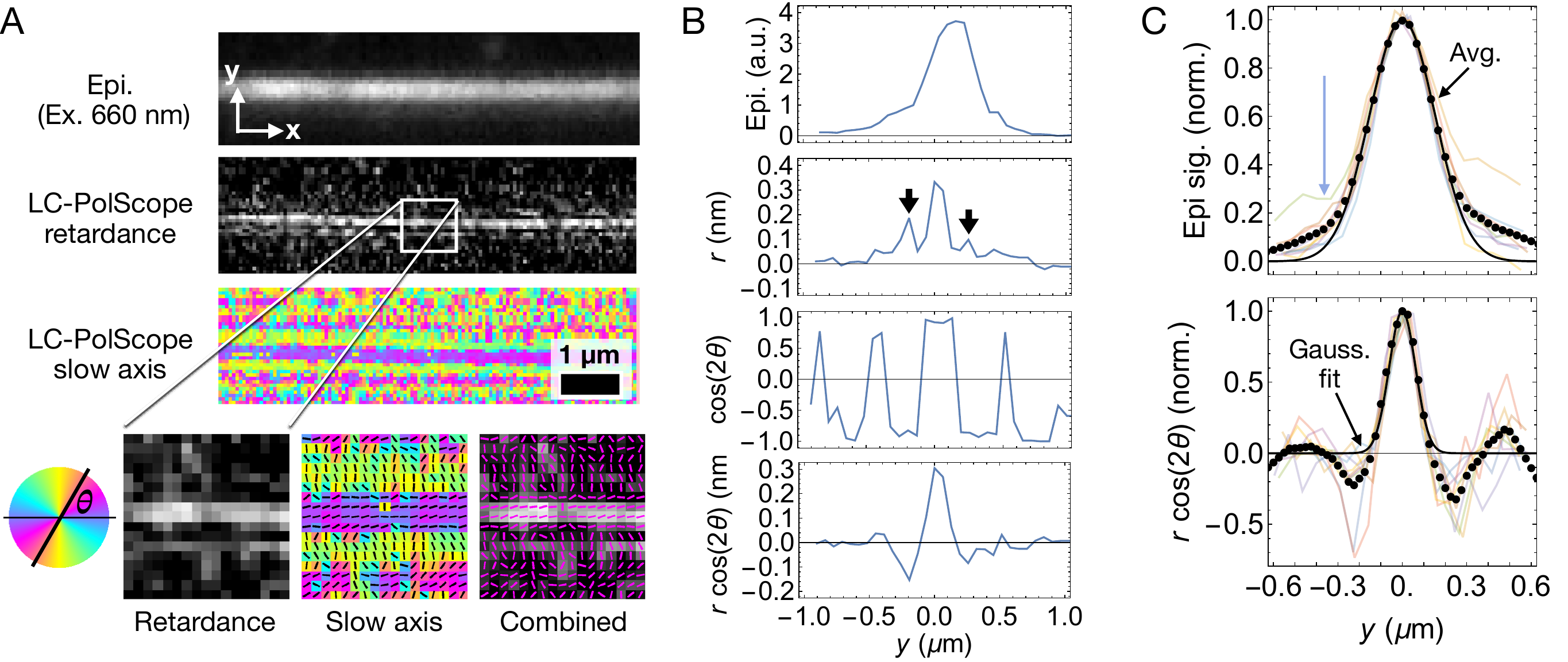}
  \caption{Characterizing the lateral optical impulse function of the LC-PolScope. (A)
\textit{Top:} Epifluorescence, retardance $r$, and slow axis $\theta$ images of a
microtubule bundle bound to a glass cover slip. \textit{Inset:} Near the central axis of
the bundle, the optical slow axis is roughly parallel to the bundle; $\sim$250 nm away
from the bundle axis, the slow axis is perpendicular to the bundle. Side length of inset
box = $1\um$. (B) Profiles of epifluorescence intensity, $r$, $\cos{(2\theta)}$, and
projected retardance $r \cos{(2 \theta)}$ in the direction $\yhat$ perpendicular to the
bundle, in a coordinate system where the origin is positioned at the center of the inset
box shown in A. From the retardance profile $r(y)$, we estimate that the bundle
contains $A_{0}^{-1}\int r(y) \text{d}y \approx 11$ microtubules. Secondary peaks in the
retardance profile are marked with black arrows. (C) \textit{Top:} Epifluorescence
profiles, centered at $y=0$ and normalized by their maximum values, for $n=9$ different
bundles, each containing between 3 and 12 microtubules (faint, continuous colored curves);
average epifluorescence profile (black dotted curve); best fit Gaussian (solid black
curve, $\sigma_{0} = 152$ nm). \textit{Bottom:} Projected retardance profiles, centered at
$y=0$ and normalized by their maximum values, for the same bundles (faint, continuous
colored curves); average normalized projected retardance profile (black dotted curve);
best fit Gaussian (solid black curve, $\sigma_{0} = 61$ nm). }
\label{sifig:polscope-resolution}
\end{figure}

We then imaged the surface-bound microtubule bundles (Fig.\ref{sifig:polscope-resolution}A). To estimate the LC-PolScope OIF from the resulting data, we follow the analysis in~\cite{oldenbourg1998birefringence}. We first used the retardance signal to estimate the number of microtubules in the bundle (Fig.\ref{sifig:polscope-resolution}) and select for further analysis bundles containing no more than 12 microtubules. Assuming microtubules in these bundles are close-packed, their diameters should be $\lesssim 2 \sqrt{3} d_{0} \approx 90$ nm, where $d_{0} \approx 25$ nm is the diameter of an individual microtubule~\cite{needleman2004higher, lemma2024structure} (Fig.\ref{sifig:polscope-resolution}B). Since the bundle diameter is significantly less than $\lambda_{\text{im}}$, the effect of the finite size of the bundle is minimized, and we may approximate the bundle as a line source of birefringence. To find the LC-PolScope OIF, we plot the projected retardance $r \cos{2 \theta}$ as a function of $y$, the direction perpendicular to the bundle axis, for $n=8$ bundles containing between 3 and 12 microtubules each (Fig.~\ref{sifig:polscope-resolution}). We next take the average of these curves and fit to a Gaussian (Fig.\ref{sifig:polscope-resolution}C). The best-fit Gaussian has a width $\sigma_{0} \approx 61$ nm, comparable to previous measurements \cite{oldenbourg1998birefringence}. This value of $\sigma_{0}$ corresponds to an optical resolution of $\approx 144$ nm, the full width at half maximum (FWHM) of the Gaussian \cite{masters2020concepts}. Finally, we compared the lateral LC-PolScope resolution to that of the far-red epifluorescence channel (excitation maximum $\lambda_{\text{ex}} \approx 660$ nm), obtained by fitting the average epifluorescence intensity along $y$. We find that the best-fit Gaussian has standard deviation $\sigma_{0}^{\text{Epi}} \approx 152$ nm, corresponding to a FWHM resolution of $\approx 358$ nm. The significantly enhanced resolution of the LC-PolScope compared to epifluorescence is presumably due to the semi-coherent nature of the data, which combines intensity (retardance) and phase (slow axis) information.

\section{Active Liquid Crystal Theory of the HeLa Spindle}
\label{sisec:model}
\subsection{Full Equations of Motion, Including Noise Terms}
The deterministic terms in our coarse-grained model are summarized by Eqns.~\eqref{eq:mt_density} and \eqref{eq:mt_director} in the Main Text. In terms of the normalized concentration $C = \rho/\rho_{0} = \Theta \rho/\Gamma_{0}$, the complete model, including independent Gaussian-distributed noise in both fields, may be written
% \begin{align}
% \frac{\partial C}{\partial t}  &= \Theta (1-C) - \nabla \cdot \left[ - D \nabla C + v_{1} C \Nhat \right] + \SC(\Rvec,t); \label{eq:si_conc}\\
% \frac{\partial N_\alpha}{\partial t} &= K \nabla^{2} N_{\alpha}+ \SN_{\alpha}(\Rvec,t), \label{eq:si_director}
% \end{align}
\begin{align}
\frac{\partial C}{\partial t}  &= \Theta (1-C) - \nabla \cdot \left[ - D \nabla C + v_{1} C \Nhat \right] + \SC(\Rvec,t); \label{eq:si_conc}\\
\frac{\partial N_\alpha}{\partial t} &= \left( \delta_{\alpha \beta} - N_\alpha N_\beta \right)[K \nabla^{2} N_{\beta}+ \SN_{\beta}(\Rvec,t)], \label{eq:si_director}
\end{align}

where the noise terms are fully characterized by the equations
\begin{gather*}
  \langle \SC (\Rvec,t) \rangle = 0 ;\qquad \langle \SC(\Rvec,t) \SC(\Rvec',t') \rangle = (\SCZero)^{2} \delta_{\alpha \beta}\delta(\Rvec-\Rvec') \delta(t-t');\\
\langle \SN_{\alpha}(\Rvec,t) \rangle = 0 ;\qquad \langle \SN_{\alpha}(\Rvec,t) \SN_{\beta}(\Rvec',t') \rangle = (\SNZero)^{2} \delta_{\alpha \beta}\delta(\Rvec-\Rvec') \delta(t-t');
\end{gather*}
where $\alpha, \beta \in \{y,z \}$, $\delta_{\alpha \beta}$ is the Kronecker delta, and $\delta(\cdot)$ is the Dirac delta. Noise amplitudes are given by the parameters $\SCZero$ and $\SNZero$. These equations of motion are equivalent to those derived in reference \cite{brugues2014physical}, with one exception: whereas that reference includes noise in the director equation only, Eqns.~\eqref{eq:si_conc} and \eqref{eq:si_director} feature independent noise in both equations.

\subsection{Numerical Estimation of Steady-State Director Field}
\label{sisec:numerical-solution}
\subsubsection{Steady-State Equation for Director Angle in Cylindrical Coordinates}
In order to compare the predictions of our model with the rotation-averaged or time-averaged orientation fields we measured, we numerically solved Eqn.~\ref{eq:NhatSS} in the Main Text, supplemented with boundary conditions appropriate for HeLa spindles. Under the assumption of cylindrical symmetry in the $\xhat$ axis, Eqn.~\ref{eq:NhatSS} in may be written in terms of the polar and azimuthal fields in the $\Phi=0$ plane, $\theta(x,s)$ and $\phi(x,s)$  ~\cite{press1975elastic}. In cylindrical coordinates $(x, s, \Phi )$,
\begin{equation}
\label{sieq:thetaanasatz}
\Nhat(x,s) = \cos{\theta (x,s)} \xhat + \sin{\theta (x,s)} \cos{\phi(x,s)}\shat + \sin{\theta (x,s)} \sin{\phi(x,s)}\Phihat,
\end{equation}
where $s = \sqrt{y^{2}+z^{2}}$ and $\Phi = \tan^{-1}(y/z)$ are the radial and azimuthal coordinates. This parameterization enforces the unit-length constraint \( |\widehat{\mathbf{N}}| = 1 \). The governing equations for \(\theta(x, s) \) and \( \phi(x, s) \) are obtained by minimizing the Frank elastic free energy under the one-constant approximation, in which all elastic moduli are set equal. This leads to the vectorial Laplace equation \( (\mathbf{I} - \Nhat \Nhat )\nabla^2 \widehat{\mathbf{N}} = 0 \) for the director field. Substituting the angular ansatz (Eqn.~\ref{sieq:thetaanasatz}) into this equation yields two coupled nonlinear partial differential equations for  \(\theta(x, s) \) and \( \phi(x, s) \). In the twist-free case (\( \phi = 0 \)), the director field lies entirely in the \( x \)-\( s \) plane and the system reduces to a single scalar equation for \( \theta(x, s) \) ~\cite{press1975elastic}:
% Then, the 3D vector Laplace equation, together with the normalization condition $| \Nhat| = 1$, reduces to two 2D scalar partial differential equations for $\theta(x,s)$ and $\phi(x,s)$. If we further assume that $\Nhat$ is twist-free ($\phi = 0$), the polar angle field $\theta(x,s)$ fully defines $\Nhat$. It is governed by the equation ~\cite{press1975elastic}
\begin{equation}
\label{sieq:thetaEqn}
\theta_{ss} +\frac{1}{s} \theta_{s} + \theta_{xx} = \frac{\sin{\theta}\cos{\theta}}{s^{2}}.
\end{equation}

\subsubsection{Numerical Solution}
Using the Mathematica software package, we numerically solved Eqn.~\ref{sieq:thetaEqn}, supplemented with appropriate boundary conditions. Key details include:
%Complete codes are available at \cite{HeLaZenodoDatabase}
\begin{itemize}
\item For electron tomography data, Eqn.~(\ref{sieq:thetaEqn}) is solved separately for each spindle, using boundary conditions defined by the spindle's best-fit ellipsoidal boundary and the experimentally measured centriole positions $(\pm d$,0,0).
\item For LC-PolScope data, Eqn.~(\ref{sieq:thetaEqn}) is solved separately for each spindle, using boundary conditions defined by the spindle's best-fit ellipsoidal boundary; the interpolar half-distance $d$ is treated as a fit parameter. The best-fit value of $d$ is chosen by comparing the predicted orientation field with the experimentally measured slow axis field.
\item At the spindle boundary, the director lies tangent to the boundary.
\item The poles (centrioles) are modeled as $+1$ defects. This is implemented numerically by forcing the nematic field to lie perpendicular to the surface of two disks centered at $(\pm d,0 )$ in the $x-s$ plane. The disks are given a nominal radius of 0.025 $a$, where $a$ is the long semi-axis of the ellipsoidal boundary.
\item To satisfy the requirements of topology, we place additional -1/2 defects at $(\pm a,0)$, i.e.~at the left and right extrema of the boundary ellipse. These features are implemented by smoothly switching the boundary condition in the neighborhoods of the extrema of the ellipse, such that the director lies perpendicular to the boundary at $(\pm a,0)$. 
\item We numerically solve Eqn.~(\ref{sieq:thetaEqn}) using the in-built Mathematica solver NDSolveValue[]. We use a custom-written adaptive grid generator that refines the mesh in the vicinity of the topological defects. The solution is found in the upper right quadrant ($x>0$, $s>0$), and mirrored through the $x-$ and $s-$axes to generate a full solution. Our results are robust to the details of adaptive grid generation, choice of the numerical integrator, and topological defect enforcement.
\end{itemize}

\subsection{Equations of Motion for Small Fluctuations About Steady State}
To test whether Eqns.~\eqref{eq:si_conc} and \eqref{eq:si_director} accurately describe the patterns of fluctuations observed in the electron tomography and LC-PolScope Data, we calculate the forms of these correlations predicted by the equations. We analyze density fluctuations around a uniform steady state in both concentration and orientation,
\begin{equation*}
C = 1 + \delta C(\Rvec,t); \qquad \Nhat = \xhat + \dN(\Rvec, t).
\end{equation*}
In the analysis boxes defined in the Main Text (Fig.~1), we find that the average deviation from uniform concentration is $\lesssim 20\%$; the average deviation of the predicted director from horizontal is $\lesssim 0.2$ rad; we therefore assume that the assumption of uniform steady-state is reasonable. To analyze the relaxation of orientation fluctuations around a strongly aligned nematic state, we expand the unit vector field as
\[
\mathbf{N} \approx
\begin{bmatrix}
1 - \tfrac{1}{2}(\delta N_y^2 + \delta N_z^2) \\
\delta N_y \\
\delta N_z
\end{bmatrix},
\quad \text{with } \delta N_y,\, \delta N_z \ll 1.
\]
Substituting this ansatz into the projected evolution equation and retaining leading-order terms, we obtain equations for the transverse components:
\begin{equation}\label{eq:si_directorFluc}
    \frac{\partial \delta N_\alpha}{\partial t} \approx K \nabla^2 \delta N_\alpha + S_\alpha^{{N}},
\quad \text{for } \alpha = y, z. 
\end{equation}
Similarly, we can derive the equation of motion of the microtubule density fluctuations $\delta C$,
\begin{equation}\label{eq:si_concFluc}
      \frac{\partial \delta C}{\partial t}  = -\Theta \delta C + D \nabla^{2} \delta C - v_{1} \nabla \cdot [ \dN + \delta C \xhat ] + \SC(\Rvec,t) 
\end{equation}

\subsection{Fourier-Space Equations of Motion; (3+1)D Correlation Functions}
Taking the Fourier transform of Eqns.~(\ref{eq:si_concFluc}) and (\ref{eq:si_directorFluc}), we obtain
\begin{gather*}
-i \w \dCT = -\Theta \dCT -  DQ^{2}\dCT + i v_{1} [ q_{x} \dCT + q_{y} \dNyT + q_{z} \dNzT ] + \SCT (\mathbf{Q}, \w);\\
-i \w \dNyT = -K Q^{2}\dNyT + \SNyT (\mathbf{Q}, \w);\qquad
 -i \w \dNzT = -K Q^{2}\dNzT + \SNzT  (\mathbf{Q}, \w).
\end{gather*}
Here, $\delta \widetilde{N}_\alpha(\mathbf{Q}, \w)$ is the Fourier transform of $\delta N_\alpha(\mathbf{R}, t)$ in space and time, and $\tilde{S}_y^n(\w, \mathbf{Q} )$ is the Fourier transform of $S_y^n(\mathbf{R}, t)$ for $\alpha \in \{ y,z \}$. The Fourier-transformed noise fields satisfy
\begin{gather*}
  \langle \SCT (\Qvec_{1},t) \rangle = 0 ;\qquad \langle \SCT(\Qvec_{1},t) \SCT(\Qvec_{2},t') \rangle = (\SCZero)^{2} \delta(\Qvec_{1}+\Qvec_{2}) \delta(\w_{1}+\w_{2});\\
\langle \SNT_{\alpha}(\Qvec_{1},t) \rangle = 0 ;\qquad \langle \SNT_{\alpha}(\Qvec_{1},t) \SNT_{\beta}(\Qvec_{2},t') \rangle = (\SNZero)^{2} \delta_{\alpha \beta}\delta(\Qvec_{1}+\Qvec_{2}) \delta(\w_{1}+\w_{2});
\end{gather*}

These equations may be solved to find the Fourier-transformed fluctuations,
\begin{equation*}
  \dCT =  \frac{ i v_1 \color{black} (q_{y} \dNyT + q_{z} \dNzT) + \SCT (\mathbf{Q}, \w)}{-i(\omega + v_{1} q_{x}) + \color{black}\Theta +  DQ^{2}  };\qquad
\dNyT = \frac{\SNyT (\mathbf{Q}, \w)}{-i \w + K Q^{2}};\qquad  \dNzT = \frac{\SNzT (\mathbf{Q}, \w)}{-i \w + K Q^{2}},
\end{equation*}

from which we obtain the spatiotemporal correlation functions $\CCC(\Qvec, \w)$,  $\CNN(\Qvec, \w)$, and $\CNC(\Qvec, \w)$,
\begin{gather}
  \CCC(\Qvec, \w) = \frac{1}{V \tau_{0}}\dCT(\Qvec,\w) \dCT(-\Qvec,-\w) =
  \frac{(q_y^2+q_z^2) v_1^{2} (\SNZero)^2 + (K^2Q^4+\w^2) (\SCZero)^2}{(K^2Q^4+\w^2)\left((q_x v_1+\w)^2+(D Q^{2}+\Theta)^{2} \right)}; \nonumber \\
  \CNN(\Qvec, \w)  =  \frac{1}{V \tau_{0}} \dNyT(\Qvec,\w) \dNyT (-\Qvec,-\w)  = \frac{1}{V \tau_{0}} \dNzT(\Qvec,\w) \dNzT (-\Qvec,-\w) = \frac{(\SNZero)^{2}}{\w^{2} + K^{2} Q^{4}};\nonumber\\
  \CNC(\Qvec, \w)  =  \frac{1}{V \tau_{0}} \dNyT(\Qvec,\w) \dCT (-\Qvec,-\w) = -\frac{iq_y v_1 (\SNZero)^2}{(K^2Q^4+\w^2)( (D Q^{2}+\Theta)+i(q_xv_1+\w))},
  \label{sieq:Cs3D}
  \end{gather}
  where $V$ is the analysis volume and $\tau_{0}$ is the total time over with the signal was recorded. The equal-time correlation functions $\SCC(\Qvec)$,  $\SNN(\Qvec)$, and $\SNC(\Qvec, \w)$ are obtained by integrating over the frequency domain,
  \begin{gather}
    \SCC(\Qvec) = \frac{1}{2 \pi }\int \CCC(\Qvec, \w) \dd \w =\frac{(\SCZero)^{2}}{2(D Q^{2}+\Theta)}
+ \frac{(q_{y}^{2}+q_{z}^{2}) (\SNZero)^{2}v_{1}^{2}((D+K)Q^{2}+\Theta)}
     {2KQ^{2}(DQ^{2}+\Theta)\bigl(q_{x}^{2}v_{1}^{2}+((D+K)Q^{2}+\Theta)^{2}\bigr)}; \nonumber \\
     \SNN(\Qvec)  =   \frac{1}{2 \pi }\int \CNN(\Qvec, \w)  \dd \w = \frac{(\SCZero)^{2}}{2 K Q^{2}}; \nonumber \\
     \SNC(\Qvec)  =  \frac{1}{2 \pi }\int \CNC(\Qvec, \w) \dd \w =- \frac{i q_y v_1 (\SNZero)^{2}}{2 K Q^{2} ((D+K)Q^{2}+ i q_{x} v_{1} + \Theta) }.
       \label{sieq:Ss3D}
\end{gather}

\subsection{Correlation Functions for $\zhat$-Projected Fields}
  \label{sec:2d_3d}
Unlike electron tomography, the LC-PolScope does not allow measurement of the full 3D director field $\Nhat(\Rvec)$. Rather, it measures the orientation field projected over the optical axis $\zhat$ (Main Text Eqns.~(\ref{eq:retard}) and (\ref{eq:polscope_2d})).
According to the projection-slice theorem of Fourier analysis, this is equivalent to taking a long wavelength limit in the $\zhat$
direction, i.e. $q_{z} \rightarrow 0$~\cite{bracewell1990numerical}. In operator notation,
\begin{equation}
\label{sqeq:proj-slice}
S_{q_{z}=0} FT_{xyz} =   FT_{xy} P_{z},
  \end{equation}
where $S_{q_{z}=0}$ is the slice operator that sets $q_{z} \rightarrow 0$ in its argument
function, $FT_{xy}$ and $FT_{xyz}$ are 2D and 3D spatial Fourier transforms, $P_{z}$ is the
projection operator over $z$, $P_{z}\{ f(x,y,z,t)\} = \int f(x,y,z,t) \dd z $. With this notation,
\begin{equation*}
\dNyT(q_{x},q_{y}, q_{z} \rightarrow 0, \w) =  FT_{xy} P_{z} \dNy \approx FT_{xy}\int_{L_{z}} \dNy
\dd z \approx L_{z} \dnyT(q_{x},q_{y},\w),
\end{equation*}
where $L_{z}$ is the sample thickness along $\zhat$. In the above expression, the first
approximation enters because of the finite sample thickness, and the second arises because
$\hat{n}$ is a normalized projection rather than an absolute projection, see Eqns.~1 of the
Main Text. (The second relation is accurate to first order in $\dN$, however.) Applying
the limit $q_{z} \rightarrow 0 $ to Eqns.~\ref{sieq:Cs3D} and \ref{sieq:Cs3D} yields
\begin{equation*}
 \label{eq:corrFuncs2D}
\cnn (\qvec, \omega) \equiv \frac{1}{V_{2+1}} \dnyT(\qvec,\omega)^{*}\dnyT(\qvec, \omega)  = \frac{(\snZero)^{2}
}{\omega^{2}  + K^{2} q^{4}}; \qquad
  \snn (\qvec) = \frac{(\snZero)^{2}}{2 K q^{2}},
\end{equation*}
where $V_{2+1} = V_{3+1}/L_{z}$ is the (2+1)D system volume, and $(\snZero)^{2} \equiv (\SNZero)^{2}/L_{z}$.

To obtain an equivalent expression for retardance fluctuations, we use equation Main Text Eqn.~\ref{eq:retard}, and make the additional assumption that the time-averaged retardance field is constant within a sample of constant thickness $L_{z}$, $\langle r(\rvec, t)\rangle_{t}  = r_{0}\approx A_{0} \rho_{0} \lambda L_{z} $, where the definitions of $A_{0}$, $\rho_{0}$, and $\lambda $ are given in the Main Text. We can express normalized retardance fluctuations as follows,
\begin{equation*}
  \delta c (\rvec,t)  \equiv \frac{\delta r(\rvec,t)}{r_{0}} = \frac{A_{0}}{r_{0}}\int_{L_{z}} \delta \rho_{2D}(\Rvec,t) dz = \frac{A_{0} \lambda}{r_{0}} \int_{L_{z}} \delta C(\Rvec,t) dz = \frac{1}{L_{z}}\int_{L_{z}} \delta C(\Rvec,t) dz.
  \end{equation*}
  We can then apply the projection-slice theorem to obtain
  \begin{equation*}
  \dCT(q_{x},q_{y},q_{z}\to 0,\omega) = L_{z} \dcT (\qvec,\w),  
  \end{equation*}
  and thus the correlation functions given in the Main Text,
  \begin{gather*}
    \ccc(\qvec,\w)  = \frac{q_y^2 v_1^{2} (\snZero)^2 + (K^2q^4+\w^2) (\scZero)^2}{(K^2 q^4+\w^2)\left((q_{x} v_{1}+\w)^2+(D q^{2}+\Theta)^{2}\right)}; \\
    \scc(\qvec) = \frac{(\scZero)^{2}}{2(D q^{2}+\Theta)}
+ \frac{q_{y}^{2}(\snZero)^{2}v_{1}^{2}((D+K)q^{2}+\Theta)}
     {2Kq^{2}(D  q^{2}+\Theta)\bigl(q_{x}^{2}v_{1}^{2}+((D+K)q^{2}+\Theta)^{2}\bigr)}; \\
\snc(\qvec)  = \frac{i q_y v_1 (\snZero)^{2}}{2 K q^{2} ((D+K)q^{2}+ i q_{x} v_{1} + \Theta) },
\end{gather*}
where $(\scZero)^{2} \equiv (\SCZero)^{2}/L_{z}$.

\subsection{Constant-$x$ Slice Correlation Function $\RCC(q_{r})$}
  The constant-$x$ slice correlation function $\RCC(q_{r})$ is obtained by further marginalizing $\SCC(\Qvec) $ over the spindle axis direction $\xhat$,
 \begin{equation*}
    \RCC(\qvec_{\perp}) = \frac{1}{2 \pi }\int \SCC(\Qvec) \dd q_{x}.
  \end{equation*}
Substituting the explicit form of the structure factor $S_{cc}(\mathbf{Q})$, we get
\begin{equation*}
    \RCC(\Qvec) = \frac{1}{2\pi }\int \left(\frac{(\SCZero)^{2}}{2(D Q^{2}+\Theta)}
+ \frac{(q_{y}^{2}+q_{z}^{2}) (\SNZero)^{2}v_{1}^{2}((D+K)Q^{2}+\Theta)}
     {2KQ^{2}(DQ^{2}+\Theta)\bigl(q_{x}^{2}v_{1}^{2}+((D+K)Q^{2}+\Theta)^{2}\bigr)} \right) dq_x.
\end{equation*}
The second term in the integrand can be decomposed into analytically integrable components, leading to the expression:
\begin{equation*}
   \RCC(\Qvec)  = \frac{L_z (\scZero)^2}{4 \sqrt{D} \sqrt{D q_\perp^2 + \Theta}} + \frac{L_z (\snZero)^{2}}{2\pi} \int (T_1(q_x, \mathbf{q}_\perp) + T_2(q_x, \mathbf{q}_\perp))dq_x,
\end{equation*}
where $\scZero$ and $\snZero$ are the projected variables as described in the previous section. The functions $T_1(q_x, q_\perp)$ and $T_2(q_x, q_\perp)$ are described as follows:
\begin{gather}
        T_1(q_x, \mathbf{q}_\perp)  
          =  \frac{ q_s''^2 }
     {2 D (D+K)^2 (q_x^2 + e_1^2)\bigl( q_{x}^{2} w^2 +(q_x^2 + q_s'^2  )^{2}\bigr)},  \nonumber \\
        T_2(q_x, \mathbf{q}_\perp)    
          =  \frac{ q_s''^2 }
     {2 K (D+K)^2 (q_x^2 + e_2^2)\bigl( q_{x}^{2} w^2 +(q_x^2 + q_s'^2  )^{2}\bigr)}, \nonumber
\end{gather}
where $q_s = \vert \mathbf{q}_\perp \vert; K'= D+K; \: R = \sqrt{4q_{s'}^2 + w^2}; \:
\alpha_{\pm} = \sqrt{2q_{s'}^2 + w(w \pm R)};\:
\alpha = \alpha_{+} + \alpha_{-}; \:
\alpha'(q_s) = w \left( R (\alpha_{-} - \alpha_{+}) + w\alpha \right); \:
q_{s'}^2 = q_s^{2} + \frac{\Theta}{D + K}; \:
w^2 = \frac{v_1^2}{(D + K)^2};\:
e_1^2 = q_s^2 + \frac{\Theta}{D}; \:
e_2^2 = q_s^2; \:
q_{s''}^2 = q_s^2 v_1^2$. Evaluating this integral, and using the fact that that the predicted correlation function is isotropic in the plane perpendicular to the $\xhat$-axis,  $\RCC(\qvec_{\perp}) = \RCC( q_{s})$, where $q_{s} = (q_{y}^{2}+q_{z})^{1/2}$, gives Eqn.~(\ref{eq:Rcc}) in the Main Text.

\subsubsection{Applicability of These Approximations to LC-PolScope Data}
Near the central spindle, where we calculate the projected electron tomography correlation functions, all the approximations in the above calculations should be reasonably accurate. In our computations of LC-PolScope correlation functions, however, we use an analysis box that is displaced from the central spindle (Main Text Fig.~2D). We do this to minimize artifacts associated with the presence of chromosomes at the metaphase plate. However, this choice means that the analysis box may include one pole and/or the spindle boundary. Near those features, a number of key assumptions that we used to derive the correlation functions fail. These include: 
\begin{itemize}
\item Near the poles and the spindle boundary, the director field is not well-approximated by small perturbations about the long axis direction, $\Nhat\cdot \xhat \not\approx  1 $, and the component of the fluctuation parallel to the spindle axis, $\delta n_{x}$, is comparable in magnitude to the perpendicular component $\delta n_{y}$.
\item Near the poles and the spindle boundary, the retardance formula given in the Main Text (Eqn.~\eqref{eq:retard}) is expected to fail; an expression such as Eqn.~\eqref{eq:retardFull}, that takes into account the finite angle between the director and the image plane, is more appropriate.
\item Near the boundary, the spindle thickness in the $\zhat$-direction goes to zero; this conflicts with the assumption of constant sample thickness in the calculation of the projected correlation functions.
\end{itemize}
Given all of these complications, which we either ignore or correct for in an ad-hoc manner, the excellent agreement between the correlation functions calculated from LC-PolScope data and the numerically projected correlation functions calculated from electron tomography is remarkable, and may indicate that the fluctuation signals are dominated by microtubules in that are close enough to the central-spindle that the approximations remain valid.

\subsubsection{$\zhat$-Projected Correlation Functions for Minimal Model, Without Active Polar Transport}
A minimal model for correlation functions may be written in which density and orientation dynamics are governed by independent local processes: microtubule turnover, diffusion, and nematic elasticity. This minimal theory is obtained by setting $v_{1} = 0 $ in Main Text Eqns.~\eqref{eq:FourierCFs},
 \begin{gather*}
   \ccc(\qvec,\w)  = \frac{(\scZero)^2}{\w^2+(D q^{2}+\Theta)^{2}}; \qquad
   \cnn (\qvec, \omega)  = \frac{ (\snZero)^{2} }{\omega^{2}  + K^{2} q^{4}}; \\
   \scc(\qvec) = \frac{(\scZero)^{2}}{2(D q^{2}+\Theta)}; \qquad
   \snn (\qvec) = \frac{(\snZero)^{2}}{2 K q^{2}}; \qquad \snc(\qvec)  = 0.
\end{gather*}

\clearpage
\setlength{\tabcolsep}{10pt} % Default value: 6pt
\renewcommand{\arraystretch}{1.5}
\begin{table}
\begin{tabular}{ |c|c|c| }
 \hline
 Parameter & ET & LC-PolScope \\
 \hline \hline
  Long semi-axis $a$ ($\um$) & $8.8 \pm 1.0$ & $7.1 \pm 0.3$\\ 
 \hline
 Short semi-axis $b$ ($\um$) & $5.2 \pm 0.2$ & $5.6 \pm 0.3 $ \\ 
 \hline
Interpolar half-spacing $d$ ($\um$) & $6.5 \pm 0.3$  & $6.5^{\dagger} \pm 0.7$ \\ 
 \hline
\end{tabular}
\caption{Comparison of average spindle geometries, as estimated from electron tomography (ET) and LC-PolScope data. Each parameter (except the interpolar half-spacing in the electron tomography column) is estimated by taking the mean over all spindles we analyzed; uncertainty is estimated form the standard error in the mean. $^{\dagger}$ Due to our rescaling of electron tomography coordinates (S.I.~\ref{et-rescaling}), the mean interpolar half-spacing in the electron tomography data is not estimated but is instead defined to be equal to the mean interpolar half-spacing estimated from LC-PolScope. Un-rescaled electron tomography parameters can be calculated by dividing all values by $\psi = 1.44$.}
\label{sitab:paramcomparison}
\end{table}

\clearpage

\clearpage

\clearpage

\clearpage

\clearpage

\clearpage
 \clearpage